\pdfoutput=1

\documentclass[twocolumn,showpacs]{revtex4}
\usepackage{amssymb,amsmath,graphicx}
\usepackage{color}
\begin{document}

\title{
  %
  Trigonal warping, pseudodiffusive transport, and finite-system 
  version of the Lifshitz transition in magnetoconductance
  of~bilayer-graphene Corbino disks
}

\author{Grzegorz Rut}
\affiliation{Marian Smoluchowski Institute of Physics, 
Jagiellonian University, {\L}ojasiewicza 11, PL--30348 Krak\'{o}w, Poland}
\author{Adam Rycerz}
\affiliation{Marian Smoluchowski Institute of Physics, 
Jagiellonian University, {\L}ojasiewicza 11, PL--30348 Krak\'{o}w, Poland}

\begin{abstract}
Using the transfer matrix in the angular-momentum space we investigate
the impact of trigonal warping on magnetotransport and scaling properties
of a ballistic bilayer graphene in the Corbino geometry. 
Although the conductivity at the charge-neutrality point and zero magnetic 
field exhibits a~one-parameter scaling, 
the shot-noise characteristics, quantified by the Fano factor $\mathcal{F}$ 
and the third charge-transfer cumulant $\mathcal{R}$, remain pseudodiffusive.
This shows that the pseudodiffusive transport regime in bilayer graphene
is not related to the universal value of the conductivity but can be
identified by higher charge-transfer cumulants. 
For Corbino disks with larger radii ratios the 
conductivity is suppressed by the trigonal warping, mainly because the 
symmetry reduction amplifies backscattering for normal modes corresponding 
to angular-momentum eigenvalues $\pm{}2\hbar$. 
Weak magnetic fields enhance the conductivity,  reaching the maximal 
value near the crossover field $B_L=\frac{4}{3}\sqrt{3}\,({\hbar}/{e})\,
t't_\perp\!\left[{t_0^2a(R_{\rm o}\!-\!R_{\rm i})}\right]^{-1}$, where $t_0$ 
($t_\perp$) is the nearest-neighbor intra- (inter-) layer hopping integral,
$t'$ is the skew-interlayer hopping integral, and  $R_{\rm o}$ ($R_{\rm i}$) is 
the outer (inner) disk radius. 
For magnetic fields $B\gtrsim{}B_L$ we observe quasiperiodic conductance 
oscillations characterized by the decreasing mean value 
$\langle\sigma\rangle-\sigma_0\propto{}B_L/B$, where $\sigma_0=
(8/\pi)\,e^2/h$. 
The conductivity, as well as higher charge-transfer cumulants, show beating
patterns with an envelope period proportional to $\sqrt{B/B_L}$. 
This constitutes a~qualitative difference between the high-field ($B\gg{}B_L$) 
magnetotransport in the $t'=0$ case (earlier discussed
in Ref.\ \cite{Rut14d})
and in the $t'\neq{}0$ case, 
providing a~finite-system analog of the Lifshitz transition. 
\end{abstract}

\date{January 21, 2015}
\maketitle

\section{Introduction}
Bilayer graphene (BLG) rises as a~top-tier candidate material for the upcoming carbon-based electronics \cite{Avo07,Nov12,Bri12} either due to a~tunable band gap \cite{Mac13} or due to topologically-protected quantum channels along domain walls \cite{Mar08,Zha13,LJu15}, recently proposed to host nonlocal Einstein-Podolsky-Rosen pairs \cite{Sch15}. Low-energy physics of BLG is mainly governed by microscopic parameters describing the coupling between the two layers \cite{Mac13}, some of which are still far from being precisely determined. 

In the most common Bernal stacking, the leading tight-binding parameters: the intralayer hopping integral between nearest neighbors $t_{0}=3.16\,$eV, and the nearest neighbor interlayer hopping integral $t_{\perp}=0.38\,$eV, are both link to the basic BLG band-structure characteristics, which are the Fermi velocity 
\begin{equation}
  \label{vfedef}
  v_F=  \frac{\sqrt{3}\,t_0a}{2\hbar}\approx{}10^6\,\mbox{m/s}
\end{equation}
and the electron effective mass
\begin{equation}
  \label{mefdef}
  m_{\rm eff}=  \frac{t_\perp}{2v_F^2} \approx{} 0.033\,m_e,
\end{equation}
where $a=0.246\,$nm is the lattice constant in a~single layer and $m_e$ is the free electron mass. For the next-nearest neighbor (or {\em skew}) interlayer hopping integral $t'$, the corresponding characteristic is the Lifshitz energy
\begin{equation}
  \label{elidef}
  E_L = \frac{1}{4}\,t_{\perp}\left(t'/t_0\right)^2,
\end{equation}
which can be defined as a value of the electrochemical potential below which the Fermi surface splits into a~four-element manifold. $E_L$ is difficult to be directly determined in the experiment; values of $t'$ obtained from the infrared spectroscopy covers the range from $0.10\,$eV \cite{Mal07} up to $0.38\,$eV \cite{Kuz09}. 

On the other hand, in BLG even a~tiny band structure modification near the Dirac point due to $t'\neq{}0$ may have a significant impact on the minimal conductivity \cite{Cse07a}. For $t'=0$, both the mode-matching analysis \cite{Sny07} and the Kubo formalism \cite{Cse07b} lead to 
\begin{equation}
  \label{eq:pseudo}
  \sigma_{0}=2\sigma_{\rm MLG}=\frac{8e^2}{\pi{}h},
\end{equation}
where $\sigma_{\rm MLG}$ denotes the universal conductivity of a~monolayer \cite{Kat06a,Two06,Mia07}. For $t'\neq{}0$, similar theoretical considerations show the conductivity $\sigma(L)$ is no longer universal but size-dependent, and monotonically grows with the system size $L$. Depending on the crystallographic orientation of a~sample, the conductivity approaches $(7/3)\,\sigma_{0}\leqslant\sigma(\infty)\leqslant{}3\,\sigma_{0}$ \cite{Mog09}. The transport anisotropy appears as the secondary Dirac cones, present for $t'\neq{}0$, break the rotational invariance of the dispersion relation \cite{Mac06}. In principle, the experimental study of $\sigma(L)$ for clean bilayer samples should be sufficient to determine $t'$ \cite{Rut14a}, also in the presence of small interaction-induced energy gap \cite{Rtt11,Bao12,Rut14b}. (For instance, the value of $\sigma\approx{}2.5\,\sigma_0$ reported by Mayorov {\em et al.}\ \cite{May11} coincides with the above-mentioned large-$L$ predictions.) Unfortunately, device-to-device conductance fluctuations in real disordered systems may put the effectiveness of such a procedure in question. For this reason, new phenomena, unveiling the influence of $t'$ on transport characteristics which can be measured in a single-device setup, are desired.

\begin{figure}
\centerline{\includegraphics[width=0.9\linewidth]{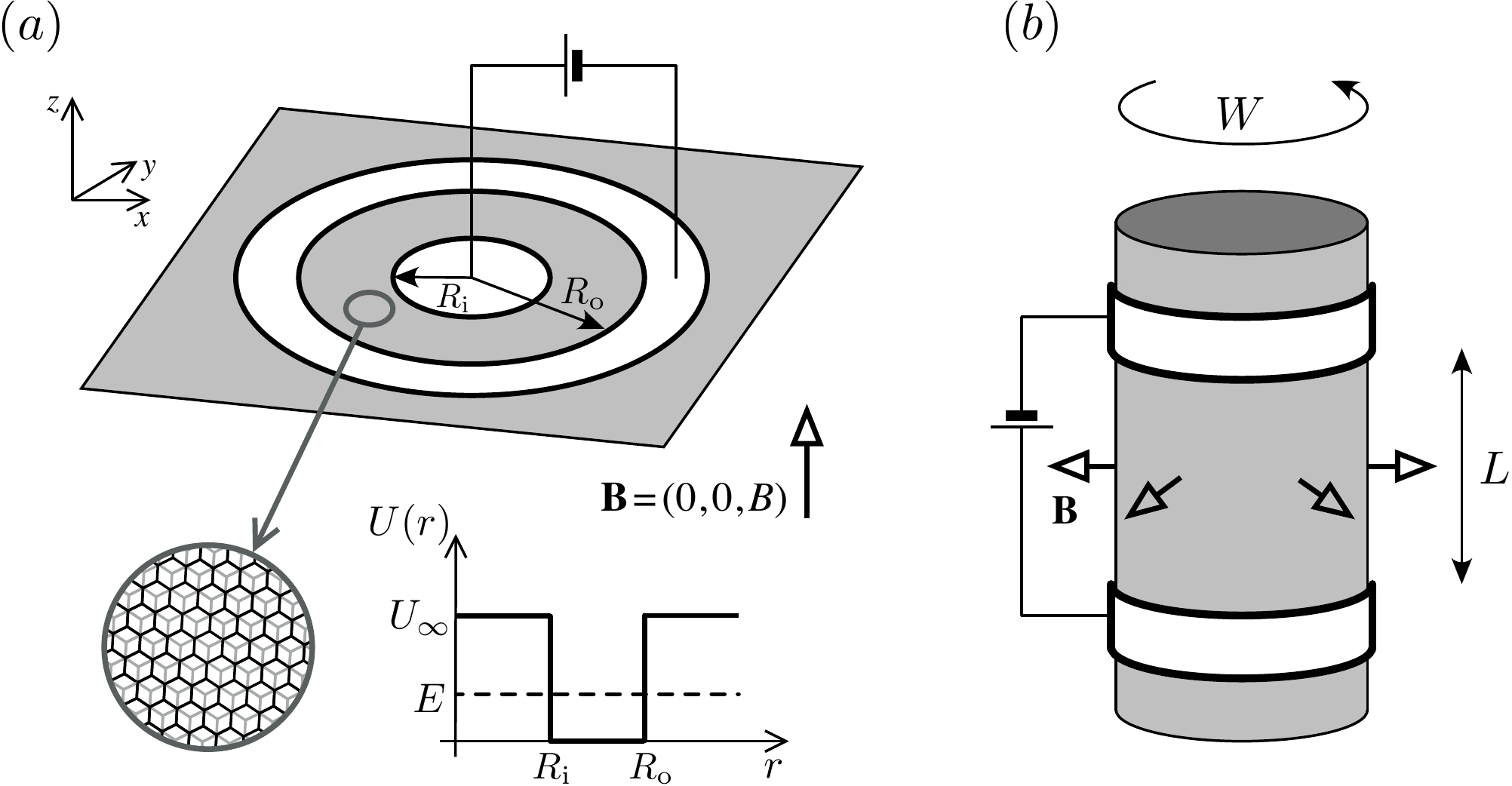}}
\caption{\label{setcoraux}
  Systems discussed in the paper (schematic). (a) The Corbino disk in Bernal-stacked bilayer graphene. The current flows through the disk-shaped area with the inner radius $R_\mathrm{i}$ and the outer radius $R_\mathrm{o}$ in a~perpendicular magnetic field $\mathbf{B}=(0,0,B)$. The coordinate system and the electrostatic potential $U(r)$ (with $r=\sqrt{x^2+y^2}$) are also shown. The leads (white areas) are modeled as infinitely-doped graphene regions ($|U_\infty|\rightarrow\infty$). (b) An artificial (nanotube-like) system formed of a BLG strip of the width $W$, contacted by two electrodes at a distance $L$ in uniform field $B$, upon applying periodic boundary conditions. 
}
\end{figure}

In this paper, we consider the Corbino geometry, in which a disk-shaped sample is surrounded from both interior and exterior sides with metallic leads [see Fig.\ \ref{setcoraux}(a)]. Such a choice is motivated by the absence of boundary effects, and irrelevance of the sample crystallographic orientation. It is worth to stress that anisotropic aspects of quantum transport are still present in such a~system as the rotational symmetry is intrinsically broken due to trigonal warping. In effect, the total angular momentum is not conserved, and the standard mode-matching method cannot be applied as in simpler systems studied earlier (namely, the monolayer disk \cite{Ryc10,Kat10} and the BLG disk with $t'=0$ \cite{Rut14d}). Here, we  overcome this difficulty by developing a~numerical transfer-matrix approach in the angular momentum space. Contrary to real-space discretization applied in some related works \cite{Two08,Ihn12,Bah13}, our approach takes benefit from the sparsity of transfer matrix in such a~representation allowing highly-efficient (albeit conceptually simple) transport calculations for physical disk diameters exceeding $10\,\mu$m. Recently developed linear-scaling approach employing the time evolution scheme \cite{Ort13} seems to be a possible counterpart but its efficient adaptation for calculating higher charge-transfer cumulants might be difficult. 

The paper is organized as follows. In Sec.\ II we present details of the system considered, the effective Dirac equation for low-energy excitations, and provide a~description of our numerical procedure. Sec.\ III focuses on the size dependence of the conductivity $\sigma$, the Fano factor ${\cal F}$, and the third charge-transfer cumulant for BLG Corbino disk at zero magnetic field. In Sec.\ IV, the effects of perpendicular magnetic field on the above-mentioned transport characteristics are discussed. Also in Sec.\ IV, the numerical results for the disk are compared with analytical ones, obtained for an artificial system [a~rectangular device with periodic boundary conditions, see Fig.\ \ref{setcoraux}(b)], for which the mode-matching analysis is possible, and the universal magnetoconductance characteristics, relevant for large system sizes and high magnetic fields, are identified. The conclusions are given in Sec.\ V.

\section{The model and the numerical approach}

\subsection{Effective Dirac equation}
The Corbino disk in Bernal-stacked BLG is depicted schematically in Fig.\ \ref{setcoraux}(a). We start our analysis from the four-band low-energy Hamiltonian for $K$ valley \cite{Mac13}, which is given by
\begin{equation}
\label{hameffu}
H=\left(\begin{array}{cccc}
0 & \pi & t_{\perp} & 0 \\
\pi^{\dagger} & 0 & 0 & \nu\pi \\
t_{\perp} & 0 & 0 & \pi^{\dagger} \\
0 & \nu\pi^{\dagger} & \pi & 0
\end{array}\right)+U(r),
\end{equation}
where $\pi=v_{F}\left(p_{x}+ip_{y}\right)=-i\hbar v_{F}e^{i\varphi}\left(\partial_{r}+i\frac{\partial_{\varphi}}{r}-\frac{eB}{2\hbar}r\right)$, with the gauge-invariant momenta $p_j=(-i\hbar\partial_j+eA_j)$ ($j=1,2$) and the symmetric gauge $(A_x,A_y)=(B/2)(-y,x)$ corresponding to the uniform magnetic field parallel to $z$-axis. We have further defined the dimensionless parameter $\nu=t'/t_{0}$, and introduced the polar coordinates $(r,\varphi)$. The potential energy $U(r)$ depends only on $r=\sqrt{x^2+y^2}$, and the remaining symbols are the same as in Eqs.\ (\ref{vfedef}--\ref{elidef}). As mentioned earlier, the available values of $t'$ following from different experimental \cite{Mal07,Kuz09,QLi09} and computational \cite{Min07} approaches are far from being consistent. Magnetotransport through BLG disk with $t'=0$ were discussed in analytical terms in Ref.\ \cite{Rut14d}. In this paper we take the values of $t'$ varying from $0.1\,$eV up to $0.3\,$eV in order to investigate numerically how it affects the system behavior. 

For the disk area, $R_{\rm i}<r<R_{\rm o}$, we set $U(r)=0$ and the effective Dirac equation $H\psi=E\psi$ (with $E$ the Fermi energy) can be written as 
\begin{equation}
\left(\begin{array}{cccc}
\epsilon & -f & -it & 0\\
-f^{*} & \epsilon & 0 & -\nu f\\
-it & 0 & \epsilon & -f^{*}\\
0 & -\nu f^{*} & -f & \epsilon
\end{array}\right)\psi\left(r,\phi\right)=0,\label{eq:Dirac}
\end{equation}
where $t=t_{\perp}/(\hbar v_{F})\equiv l_{\perp}^{-1},$ $f=e^{i\varphi}\left(\partial_{r}+i\frac{\partial_{\varphi}}{r}-\frac{1}{2l_{B}^{2}}r\right)$,
$\epsilon=E/(\hbar v_{F})$, and the magnetic length $l_{B}=\sqrt{{\hbar}/{(e|B|)}}$. 

In the absence of trigonal warping ($\nu=0$) the system possesses a~cylindrical symmetry and the effective Hamiltonian (\ref{hameffu}) commutes with the total angular momentum operator \cite{Per07b}
\begin{equation}
  \label{jztot}
  J_z=-i\hbar\partial_\varphi + \frac{\hbar}{2}
  \left(\begin{array}{cc}
      \sigma_0 & 0 \\
      0 & -\sigma_0
    \end{array}\right) + \frac{\hbar}{2}
  \left(\begin{array}{cc}
      -\sigma_z & 0 \\
      0 & \sigma_z
    \end{array}\right),
\end{equation}
where $\sigma_0$ is the $2\times{}2$ identity matrix, and $\sigma_z$ is one of the Pauli matrices. In such a case, the wavefunctions are products of angular and radial parts $\phi^{m}\left(r,\varphi\right)=e^{im\varphi}\left[\phi_{1}^{m},e^{-i\varphi}\phi_{2}^{m},\phi_{3}^{m},e^{i\varphi}\phi_{4}^{m}\right]^{T}\left(r\right)$ with $m$ being an integer angular-momentum quantum number.

\subsection{Outline of the approach}
In the presence of trigonal warping ($\nu\neq{}0$) the cylindrical symmetry is
broken, and the wavefunctions do not correspond directly to eigenstates of $J_{z}$. A generic workaround has been developed for systems with symmetry-breaking potentials (or impurities), where one can still express wavefunctions as linear combinations of eigenfunctions of an ideal system \cite{dattaArt,souma,tit}. Souma and Suzuki \cite{souma} considered quantum transport through Corbino disks in two-dimensional electron gas and showed that the effects of impurities can be studied numerically, starting from truncated wavefunctions in the basis of angular-momentum eigenstates. Here we adapt this method for BLG Corbino disk, as the term proportional to $\nu$ in the Hamiltonian (\ref{hameffu}) can be regarded as a peculiar type of a symmetry-breaking potential. 

A general solution of the Dirac equation (\ref{eq:Dirac}) can be written as an infinite linear combination of angular momentum eigenfunctions, namely
\begin{equation}
  \label{eq:cAnM}
  \psi\left(r,\phi\right)=\sum_{m}a_{m}\phi^{m}\left(r,\varphi\right),
\end{equation}
with arbitrary amplitudes $a_{m}$, $m=0,\pm{}1,\pm{}2,\dots$. Multiplying the Dirac equation (\ref{eq:Dirac}) by the factor
$e^{-il\varphi}$ (with $l$ an arbitrary integer) and integrating over the polar angle $\varphi$, we obtain the system of equations 
\begin{eqnarray}
\partial_{r}\phi_{1}^{l} & = & -g\left(l,r\right)\phi_{1}^{l}+i\epsilon\phi_{2}^{l}+i\nu t\phi_{1}^{l-3}\nonumber \\
 &  & -i\nu\epsilon\phi_{3}^{l-3}+2\nu g\left(l-2,r\right)\phi_{4}^{l-3},\nonumber \\
\partial_{r}\phi_{2}^{l} & = & i\epsilon\phi_{1}^{l}+g\left(l-1,r\right)\phi_{2}^{l}-it\phi_{3}^{l},\nonumber \\
\partial_{r}\phi_{3}^{l} & = & g\left(l,r\right)\phi_{3}^{l}+i\epsilon\phi_{4}^{l}+i\nu t\phi_{3}^{l+3}\nonumber \\
 &  & -i\nu\epsilon\phi_{1}^{l+3}-2\nu g\left(l+2,r\right)\phi_{2}^{l+3},\nonumber \\
\partial_{r}\phi_{4}^{l} & = & -it\phi_{1}^{l}+i\epsilon\phi_{3}^{l}-g\left(l+1,r\right)\phi_{4}^{l},\label{eq:uklRow}
\end{eqnarray}
where 
\begin{equation}
  \label{eq:glr}
  g\left(l,r\right)=\frac{l}{r}+\frac{r}{2l_{B}^{2}}.
\end{equation} 
Notice that the terms proportional to $\nu$ correspond to the mode mixing due to trigonal warping. The $D_{3d}$ dihedral symmetry of these terms (coinciding with the BLG lattice symmetry) results in the fact that equation for $\phi^{l}$ is coupled only to $\phi^{l-3}$ and $\phi^{l+3}$, which tremendously simplifies the numerical integration. 

Eqs.\ (\ref{eq:uklRow}) and (\ref{eq:glr}), along with the mode-matching conditions for $r=R_{\rm i}$ and  $r=R_{\rm o}$ (we model the leads as heavily-doped BLG areas), allows us to construct a transfer matrix (see Appendix~A for details) which can be utilized in the Landauer-B\"{u}ttiker formalism in order to calculate the conductivity and other charge-transfer characteristics. Typically, the convergence is reached for the wavefunction truncated by taking $|l|\lesssim{}M$ in Eq.\ (\ref{eq:uklRow}), with $M=25 - 500$ depending on the system size, the strength of the trigonal warping, and the applied field (with the upper value corresponding to $R_{\rm o}\approx{}5\,\mu$m, $t'=0.3\,$eV, and $B\approx{}80\,$T). Other computational aspects are also described in Appendix~A.

\section{Quantum transport dependence on the system size}

\subsection{One-parameter scaling}
A particularly intriguing property, arising from the earlier theoretical study of ballistic transport in BLG with skew interlayer hoppings ($t'\neq{}0$), is the one-parameter scaling \cite{Rut14b}. In the absence of disorder and electron-electron interactions the scaling function 
\begin{equation}
  \label{betadef}
  \beta(\sigma)=\frac{d\ln\!\sigma}{d\ln\!L},
\end{equation}
which plays a~central role in conceptual understanding of the metal insulator transition \cite{Efe97}, reproduces the scenario predicted for disordered Dirac systems with Coulomb interaction \cite{Ost10}. Here, the discussion is complemented by calculating the Fano factor $\mathcal{F}$ quantifying the shot-noise power, and the factor $\mathcal{R}$ quantifying the third charge transfer cumulant. 

Employing the Landauer-B\"{u}ttiker formula for the linear-response regime one can write \cite{Naz}
\begin{eqnarray}
  \sigma & = & 
  g_0\Theta_\alpha\mbox{Tr}\,\mathbf{T},
  \label{landsig} \\
  \mathcal{F} & = & 
  \frac{\mbox{Tr}\left[\,\mathbf{T}\left(\mathbf{1}-\mathbf{T}\right)\,\right]}{\mbox{Tr}\,\mathbf{T}},
  \label{landff} \\
  \mathcal{R} & = & 
  \frac{\mbox{Tr}\left[\,\mathbf{T}\left(\mathbf{1}-\mathbf{T}\right)\left(\mathbf{1}-2\mathbf{T}\right)\,\right]}{\mbox{Tr}\,\mathbf{T}},
  \label{landrr}
\end{eqnarray}
where $g_0=4e^2/h$ is the conductance quantum (with the factor $4$ following from the spin and valley degeneracies), the dimensionless prefactor ($\Theta_\alpha$) in Eq.\ (\ref{landsig}) is equal to $\Theta_{\rm Cor}=\ln(R_{\rm o}/R_{\rm i})/(2\pi)$ for the Corbino geometry of Fig.\ \ref{setcoraux}(a) or $\Theta_{\rm rec}=L/W$ for the rectangular geometry \cite{Ryc09c}, and
$\mathbf{T}=\mathbf{t}^{\dagger}\mathbf{t}$, with $\mathbf{t}$
the transmission matrix determined via the transfer matrix (see Appendix~A).
It was shown in Refs.\ \cite{Rut14a} and \cite{Rut14b} that the conductivity of ballistic BLG with $t'\neq{}0$ scales with the system size. For a~rectangular geometry, we still get $\sigma\approx{}\sigma_0$ for small systems, while for the larger ones ($L\gtrsim{}500\,$nm) the conductivity can be rationalized as
\begin{equation}
  \label{eq:asymp}
  \sigma(L)\simeq\sigma_{\star}\left[\,
    1-\left(\lambda/L\right)^{\gamma}\,\right],
\end{equation}
with typical parameters 
\begin{equation}
\label{asymparam}
  \sigma_\star\approx{}3\,\sigma_0, \ \ \ \ 
  \gamma\approx{}0.5,
\end{equation}
and $\lambda$ depending both on $t'$ and the sample orientation \cite{Rut14b}. The approximating formula given by Eqs.\ (\ref{eq:asymp}) and (\ref{asymparam}) applies for generic crystallographic orientation of the sample. However, for some particular orientations one obtains different asymptotic behaviors, including a slower power-law convergence to $\sigma_\star$ with the exponent $\gamma\approx{}0.25$ if the current is passed precisely along an armchair direction \cite{Rut14a}, or the oscillating conductivity with the lower bound of $\sigma\lesssim(7/3)\,\sigma_0$ if the current is passed precisely along a zigzag direction \cite{Mog09}. Such issues are absent in the Corbino geometry, allowing one to focus on the universal scaling properties of the material.

\begin{figure}
\includegraphics[width=0.9\linewidth]{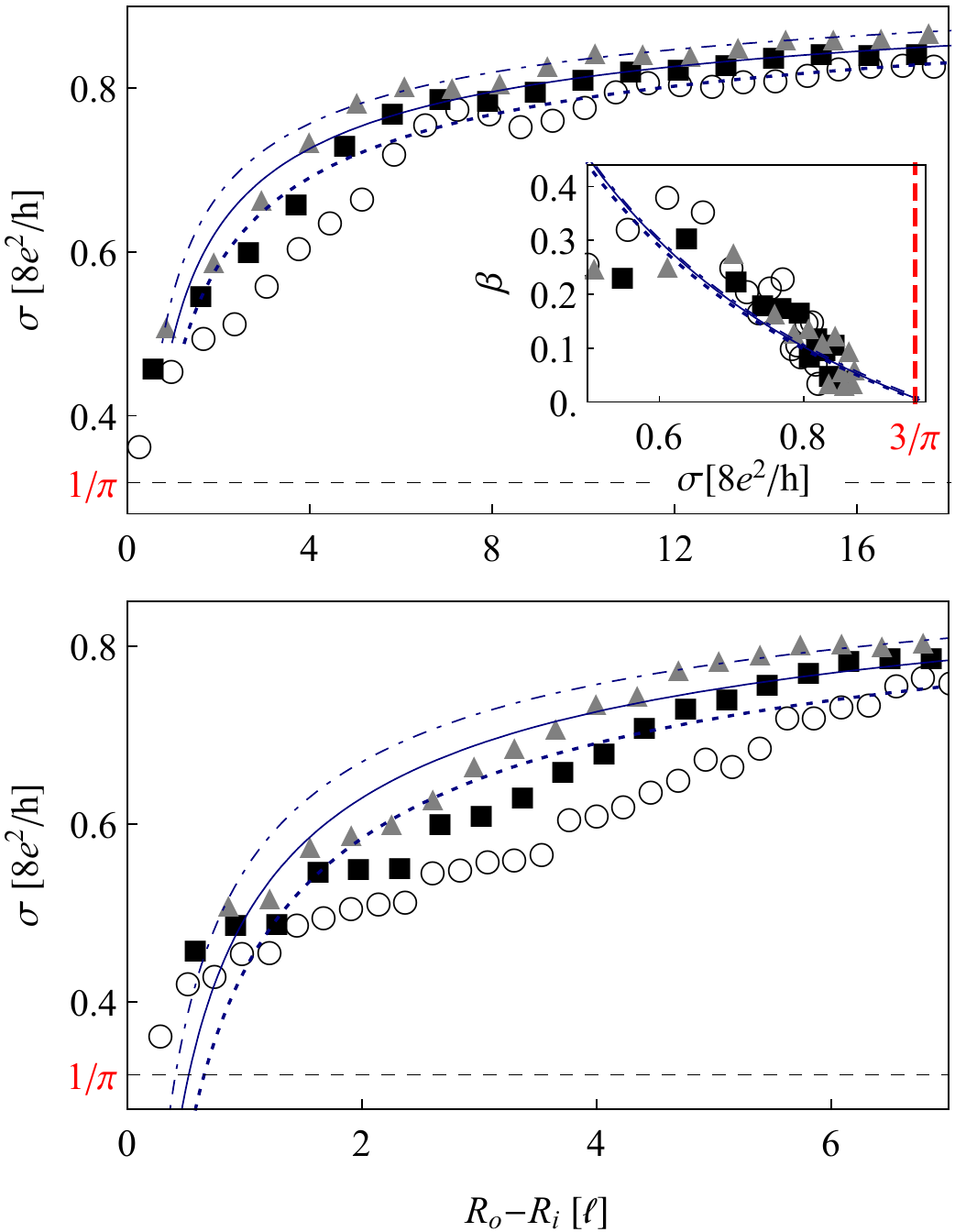}
\caption{\label{fig:Minimal-conductivity-of}
Minimal conductivity of unbiased BLG Corbino disk with the radii ratio 
$R_{\rm o}/R_{\rm i}=2$  as a~function of the radii
difference $R_{\rm o}-R_{\rm i}$, specified in the units of $\ell$ 
[see Eqs.\ (\ref{elldef},\ref{ellval}) in the main text], 
for different values of $t'$. 
Triangles, squares and circles represent the data obtained numerically 
for $t'=0.3\,$eV, $t'=0.2\,$eV, and $t'=0.1\,$eV (respectively), with the 
lines (dot-dashed, solid and dotted) depicting the approximating 
Eq.\ (\ref{eq:asymp}) with best-fitted parameters listed in Table 
\ref{tab:Comparison-of-pseudodiffusive}. 
The inset in the top panel presents the scaling 
function $\beta(\sigma)$ [see Eq.\ (\ref{betadef})], 
with $L\equiv{}R_{\rm o}-R_{\rm i}$, extracted numerically from 
$\sigma\left(R_{i}\right)$ datasets. (Notice that the best-fitted lines 
almost overlap each other.)
The bottom panel is a zoom in, for smaller radii differences, which allows
to depict the region where the actual value of the conductivity may deviate 
from Eq.\ (\ref{eq:asymp}). 
We further notice that
only selected datapoints from the bottom panel are shown in the top panel 
for clarity.  
}
\end{figure}

\begin{figure}
\includegraphics[width=0.9\linewidth]{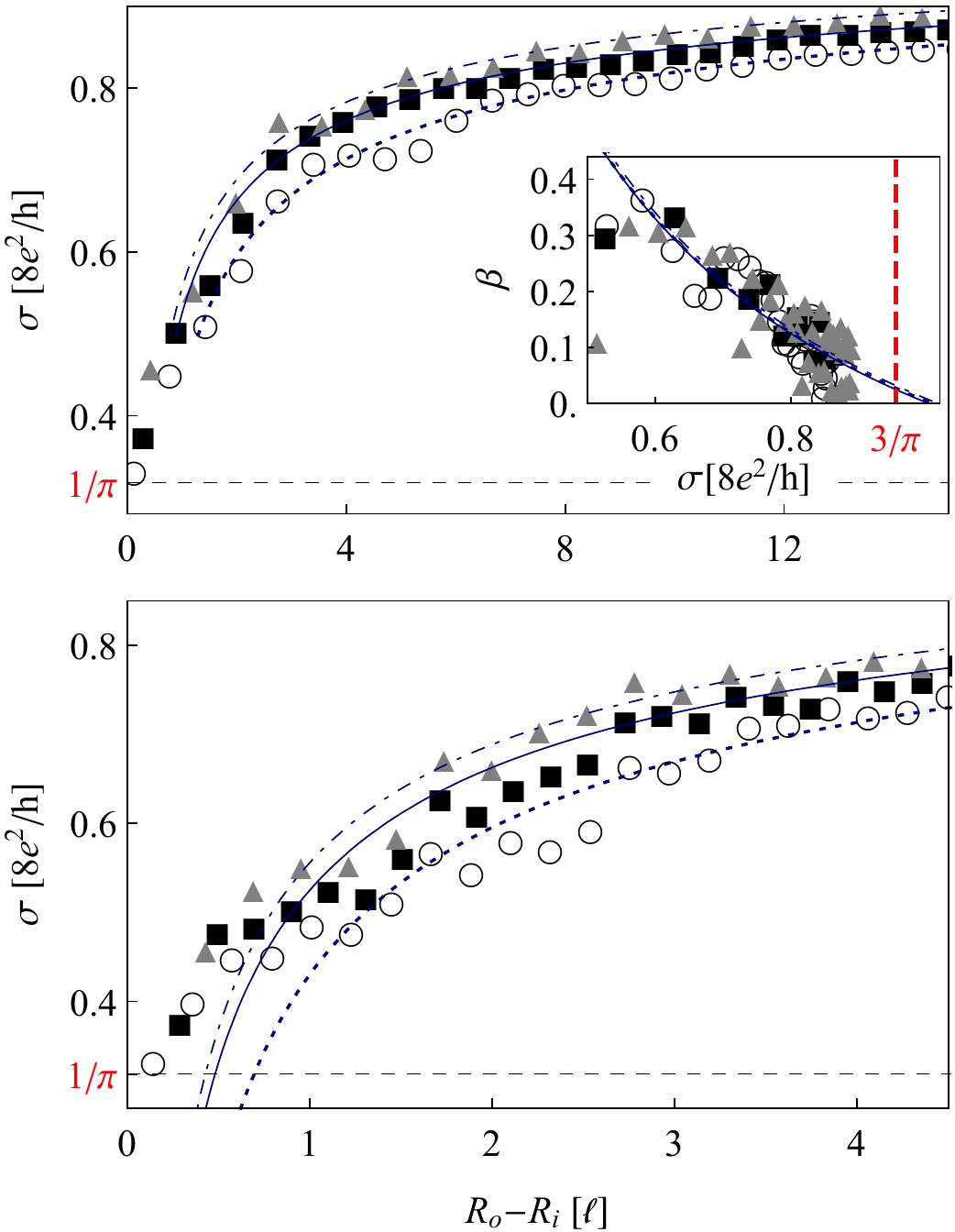}
\caption{\label{fig:Minimal-conductivity-R3o2}
Same as Fig.\ \ref{fig:Minimal-conductivity-of} but for the radii ratio
$R_{\rm o}/R_{\rm i}=1.5$. For the parameters in Eq.\ (\ref{eq:asymp}) 
corresponding to the lines depicted, see 
Table~\ref{tab:Comparison-of-pseudodiffusive-R3o2}. 
}
\end{figure}

\begin{table}[h]
\caption{
  \label{tab:Comparison-of-pseudodiffusive}
  Least-squares fitted parameters in Eq.\ (\ref{eq:asymp})
  corresponding to the lines in Fig.\ \ref{fig:Minimal-conductivity-of}. 
  The last column gives the values of 
  $L_{0.01}/\ell$, such that the function given by Eq.\ (\ref{eq:asymp})
  matches the actual conductivity with $1\%$ accuracy. 
}
\begin{tabular}{c|c|c|c|c|c}
  \hline\hline  
  $\ t'\,$(eV)$\ $ & $\ \ell\,$(nm)$\ $ &  $\ \sigma_{\star}\,(8e^{2}/h)\ $ 
  & $\ \lambda\,$(nm)$\ $ & $\ \gamma\ $ & 
  $\ L_{0.01}/\ell\ $ \\
  \hline 
  0.1  &  352  &  0.97  &  97  & $\ \ $0.47$\ \ $ &  
  6.21 \\
  0.2  &  176  &  0.97  &  40  &  0.49  &  
  5.19 \\
  0.3  &  117  &  0.98  &  21  &  0.48  &  
  4.44 \\
  \hline \hline 
\end{tabular} 
\end{table}

\begin{table}[h]
\caption{
  \label{tab:Comparison-of-pseudodiffusive-R3o2}
  Least-squares fitted parameters in Eq.\ (\ref{eq:asymp})
  corresponding to the lines in Fig.\ \ref{fig:Minimal-conductivity-R3o2}. 
  The last column is same as in Table~\ref{tab:Comparison-of-pseudodiffusive},
  but for $R_{\rm o}/R_{\rm i}=1.5$. 
}
\begin{tabular}{c|c|c|c|c|c}
  \hline\hline  
  $\ t'\,$(eV)$\ $ & $\ \ell\,$(nm)$\ $ &  $\ \sigma_{\star}\,(8e^{2}/h)\ $ 
  & $\ \lambda\,$(nm)$\ $ & $\ \gamma\ $ & 
  $\ L_{0.01}/\ell\ $ \\
  \hline 
  0.1  &  352  &  1.01  &  110  & $\ \ $0.48$\ \ $ &  
  4.79 \\
  0.2  &  176  &  1.00  &  38  &  0.49  &  
  4.29 \\
  0.3  &  117  &  1.02  &  23  &  0.49  &  
  3.79 \\
  \hline \hline 
\end{tabular} 
\end{table}

\begin{figure}
\includegraphics[width=0.9\linewidth]{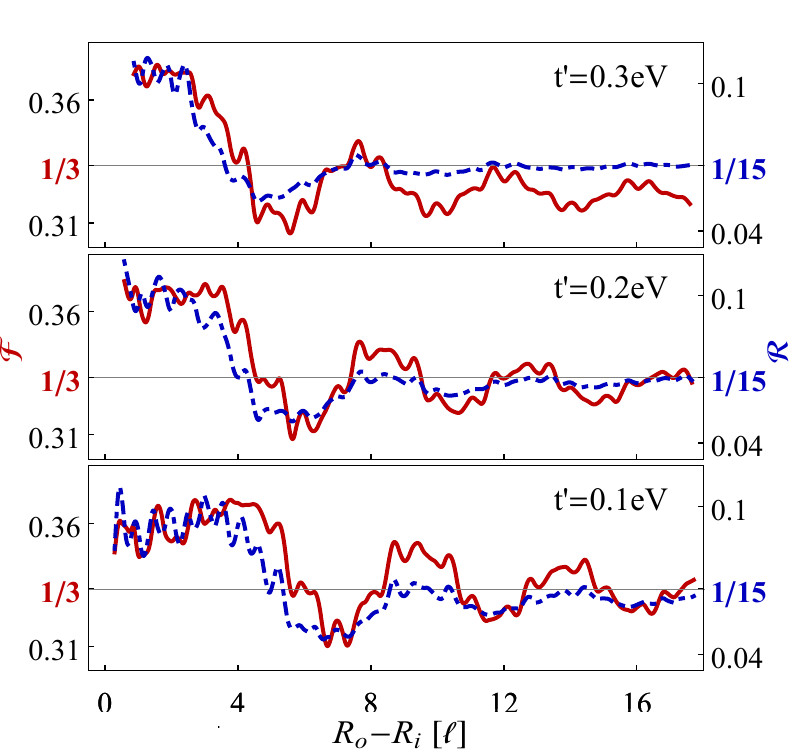}
\caption{\label{fig:Dependence-of-the}
The Fano factor $\mathcal{F}$ [red solid lines] and the third charge 
transfer cumulant $\mathcal{R}$ [blue dashed lines] for the Corbino disk with 
$R_{\rm o}/R_{\rm i}=2$ as functions of $R_{\rm o}-R_{\rm i}$. 
Skew-interlayer hopping $t'$ is varied between the panels. 
For large system size, where the Fabry-Perot oscillations become negligible, 
the numerical results obtained in the presence of trigonal warping approach 
the pseudodiffusive values of $\mathcal{F}=1/3$ and $\mathcal{R}=1/15$
[horizontal lines].
}
\end{figure}

\begin{figure}
\includegraphics[width=0.9\linewidth]{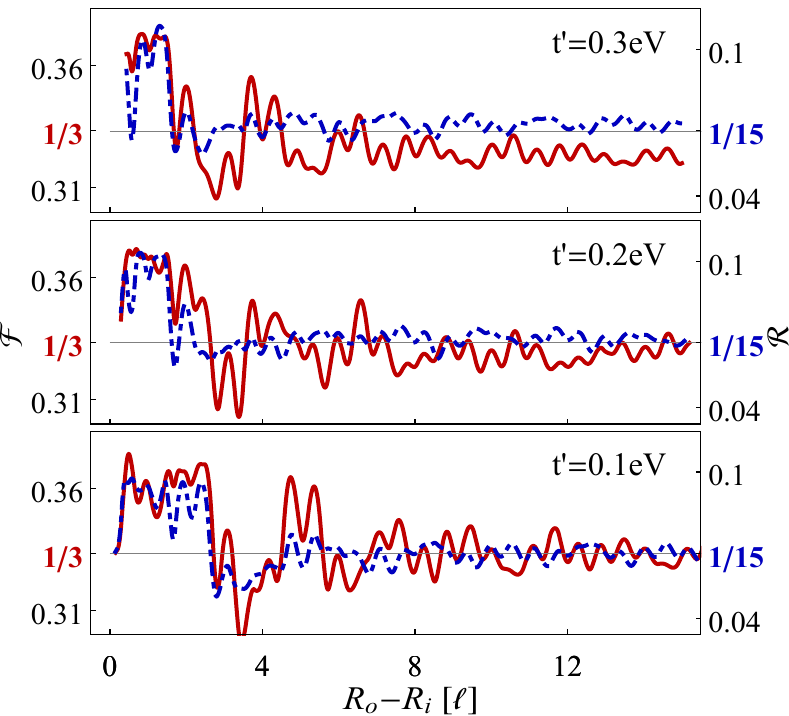}
\caption{\label{fig:Dependence-of-R3o2}
Same as Fig.\ \ref{fig:Dependence-of-the} but for $R_{\rm o}/R_{\rm i}=1.5$. 
}
\end{figure}

In the numerical calculations presented in this section, we chose radii ratios $R_{\rm o}/R_{\rm i}=2$ and  $R_{\rm o}/R_{\rm i}=1.5$, for which the number of nonzero transmission eigenvalues $T_l$ is relatively large even at the charge-neutrality point, allowing one to expect some remainders of the pseudodiffusive behavior known for a~monolayer \cite{Two06,Ryc09c}. It is also worth to mention that such radii ratios are close to that of real Corbino device in a~monolayer \cite{Pet14}. A~multimode character of the charge transport, combined with a~mode mixing due to the trigonal warping, and with a~necessity to study large systems in attempt to demonstrate a~one-parameter scaling, provides us with an excellent test case to investigate computational aspects of the numerical approach presented in Section IIB. (For the details, see Appendix~A.) 

Zero-magnetic field results are presented in 
Figs.\ \ref{fig:Minimal-conductivity-of}, \ref{fig:Minimal-conductivity-R3o2}, \ref{fig:Dependence-of-the}, and \ref{fig:Dependence-of-R3o2}.  
In order to present the data for different $t'$ on a~single plot, we have 
defined the length $\ell$, related to the distance between primary and 
secondary Dirac points in quasimomentum space
\begin{equation}
  \label{elldef}
  \frac{2\pi}{\ell}\equiv{}k_\ell = \frac{\nu}{l_\perp} 
  = \frac{2}{3}\sqrt{3}\,\frac{t't_\perp{}}{t_0^2a}, 
\end{equation}
leading to
\begin{equation}
  \label{ellval}
  \ell{}\,t'=35.2\ \text{nm}\!\cdot\!\text{eV}.
\end{equation}

Briefly speaking, the actual conductivity reaches a close-to-asymptotic behavior, described by Eq.\ (\ref{eq:asymp}) with $L\equiv{}R_{\rm o}-R_{\rm i}$, for radii differences lying in a~relatively narrow interval of $4\lesssim{}L/\ell\lesssim6$ (notice that varying the skew-interlayer hopping from $t'=0.1\,$eV to $t'=0.3\,$eV is equivalent to changing the parameter $\ell$ by a~factor of $3$ between the datasets). For larger $L$, lines in Figs.\ \ref{fig:Minimal-conductivity-of} and \ref{fig:Minimal-conductivity-R3o2}, corresponding to Eq.\ (\ref{eq:asymp}) with the least-squares fitted parameters $\sigma_\star$, $\lambda$, and $\gamma$ listed in Tables~\ref{tab:Comparison-of-pseudodiffusive} and \ref{tab:Comparison-of-pseudodiffusive-R3o2}, are closely followed by the datapoints obtained numerically for $t'=0.1\,$eV, $0.2\,$eV, and $0.3\,$eV. For each case, the specific value of $L_{0.01}$; i.e, the radii difference above which the approximating Eq.\ (\ref{eq:asymp}) matches the actual conductivity with an accuracy better than $1\%$; is also given in Table~\ref{tab:Comparison-of-pseudodiffusive} or \ref{tab:Comparison-of-pseudodiffusive-R3o2}. For smaller $L$, in particular for $L/\ell\sim{}1$ [see bottom panels in Figs.\ \ref{fig:Minimal-conductivity-of} and \ref{fig:Minimal-conductivity-R3o2}], the conductivity becomes nonuniversal (both parameter- and geometry-dependent \cite{lsmallfoo}) approaching $\sigma_0$ for $L\ll{}\ell$. 

Although the conductivity strongly deviates from the pseudodiffusive
value $\sigma_0$ (even for the lowest considered value of $t'=0.1\,$eV), 
the shot-noise power and the third charge-transfer cumulant are close to their pseudodiffusive values, i.e., $\mathcal{F}\approx{}1/3$ and $\mathcal{R}\approx{}1/15$, which are usually approached for $4\lesssim{}L/\ell\lesssim{}6$ (see Figs.\ \ref{fig:Dependence-of-the} and \ref{fig:Dependence-of-R3o2}).  

The approximating Eq.\ (\ref{eq:asymp}) leads, via Eq.\ (\ref{betadef}), to the scaling function of the form
\begin{equation}
  \beta(\sigma)\simeq{}-\gamma{}\left(1-\sigma_\star/\sigma\right). 
\end{equation}
In turn, the parametres $\gamma$ and $\sigma_{\star}$ defines the position and the slope coefficient at the attractive fixed point ($\beta(\sigma_\star)=0$, $\beta'(\sigma_\star)>0$) of the renormalization group flow \cite{Rut14b}. The scaling functions $\beta(\sigma)$ [see Eq.\ (\ref{betadef})], calculated numerically for the Corbino disks with $R_{\rm o}/R_{\rm i}=2$ and $R_{\rm o}/R_{\rm i}=1.5$ (see insets in Figs.\ \ref{fig:Minimal-conductivity-of} and \ref{fig:Minimal-conductivity-R3o2}, respectively), $t'=0.1\,$eV, $0.2\,$eV, and $0.3\,$eV, suggest that one-parameter scaling is universal with respect to the strength of the trigonal warping. These numerical results coincide with the corresponding analysis of Ref.\ \cite{Rut14b} for a rectangular sample.

\subsection{Crossover to a quantum-tunneling regime}
It was pointed out that generic MLG billiard with (at least) one narrow opening shows a~crossover to the so-called {\em quantum-tunneling} transport regime  \cite{Ryc09c}, in which charge-transfer characteristics are governed by a~limited number of quantum channels, with transmission probabilities showing a~power-law decay with the system size. In particular, for an {\em undoped} Corbino disk the conductance $G=2\pi\sigma/\ln(R_{\rm o}/R_{\rm i})$ at zero field reads
\begin{equation}
  \label{gmlgdisk}
  \frac{G_{\rm MLG}}{g_0} = \sum_{j=\pm\frac{1}{2},\pm\frac{3}{2},\dots}
  \frac{1}{\cosh^2\left[\,j\,{\ln(R_{\rm o}/R_{\rm i})}\,\right]} \simeq{} 
  \frac{8R_{\rm i}}{R_{\rm o}},
\end{equation}
where the asymptotic form applies for $R_{\rm o}\gg{}R_{\rm i}$, and represents contributions from the two channels with angular-momentum quantum numbers $j=-1/2$ and $1/2$. 

For BLG billiards, the conformal mapping technique employed in Ref.\ \cite{Ryc09c} cannot be utilized even in the absence of trigonal warping, and the existence of a quantum-tunneling regime is not obvious. For BLG Corbino disk the conductance, for $t'=0$ and at the charge-neutrality point, can be written as \cite{Rut14d}
\begin{align}
  \frac{G_{\rm BLG}(t'=0)}{g_0} &= 
  \sum_{m}\left(\,T_m^{+}+T_m^{-}\,\right), \label{gblg0dsk} \\
  T_m^{\pm} &= \frac{1}{\cosh^2\left[\,
      (m\pm{\cal A})\ln\left(R_{\rm o}/R_{\rm i}\right)\,\right]},
  \label{tmpm0dsk}
\end{align}
where the transmission probabilities $T_m^\pm$ (with $m=0,\pm{}1,\pm{}2,\dots$ being the angular-momentum quantum number) correspond to eigenvalues of the matrix ${\bf T}={\bf t}^\dagger{\bf t}$ in Eqs.\ (\ref{landsig}--\ref{landrr}), and we have further defined 
\begin{align}
  {\cal A} =& 
  -\frac{ \ln\left(\,\Upsilon-\sqrt{\Upsilon^{2}-1}\,\right)}{2\ln\left(R_{\rm o}/R_{\rm i}\right)}, 
  \label{caladef} \\
  \Upsilon =& 
  \cosh\left[\ln\left(\frac{R_{\rm o}}{R_{\rm i}}\right)\right] 
  + \frac{ R_{\rm o}^2\!-\!R_{\rm i}^2 }{ 4l_\perp^2 }
  \sinh\left[\ln\left(\frac{R_{\rm o}}{R_{\rm i}}\right)\right]. 
  \label{upsidef}
\end{align}

We focus now on the system behavior for $R_{\rm o}\gg{}R_{\rm i}\gg{}l_\perp$ \cite{lpefoo}. In such a~parameter range, Eqs.\ (\ref{caladef},$\,$\ref{upsidef}) lead to 
\begin{equation}
  \label{calaappr}
  {\cal A} \approx \frac{3}{2} + 
  \frac{\ln\left[\,R_{\rm i}/\left(2l_\perp\right)\,\right]}{\ln\left(R_{\rm o}/R_{\rm i}\right)}.
\end{equation}
For any integer ${\cal A}=q\geqslant{}2$ the conductance, analyzed as a~function of $R_{\rm o}/R_{\rm i}$, reaches a~local maximum with $G_{\rm BLG}\approx{}2g_0$, following from the presence of two ballistic channels with $T_{q}^-=T_{-q}^+=1$ [see Eqs.\ (\ref{gblg0dsk}) and (\ref{tmpm0dsk})], occurring at 
\begin{equation}
  \label{maxgblg}
  \frac{R_{\rm o}}{R_{\rm i}} \approx 
  \left( \frac{R_{\rm i} }{ 2l_\perp } \right) ^{2/(2q-3)}. 
\end{equation}
Similarly, Eq.\ (\ref{maxgblg}) with half-odd integer $q\geqslant{}5/2$ approximates a~local conductance minimum (as ${\cal A}\approx{}q$) with $G_{\rm BLG}\approx{}16g_0R_{\rm i}/R_{\rm o}$; i.e., {\em twice} as large as the MLG disk conductance, see Eq.\ (\ref{gmlgdisk}); and four dominant channels characterized by $T_{q+1/2}^-=T_{-q-1/2}^+\approx{}T_{q-1/2}^-=T_{-q+1/2}^+\approx{}4R_{\rm i}/R_{\rm o}$. In turn, a~quantum-tunneling regime reappears periodically when varying $R_{\rm o}/R_{\rm i}$, near any local conductance minimum. The number of well-pronounced minima can roughly be estimated as $0.72\times\ln[\,R_{\rm i}/(2l_\perp)\,]$, as for $R_{\rm o}/R_{\rm i}\lesssim{}4$ the system enters a~multimode pseudodiffusive transport regime. On the other hand, for $R_{\rm o}/R_{\rm i}\gtrsim{}R_{\rm i}^2/(4\l_\perp^2)$, where the threshold value corresponds to the last conductance maximum following from Eq.\ (\ref{maxgblg}) with $q=2$, charge transport is governed by two equivalent channels, with angular momenta $\pm{}2\hbar$ and $T_2^-=T_{-2}^+$ monotonically decaying with increasing $R_{\rm o}/R_{\rm i}$. This suggests the system may re-enter a~quantum-tunneling limit for  $R_{\rm o}/R_{\rm i}\gg{}R_{\rm i}^2/(4\l_\perp^2)$. 

In fact, for a~fixed $R_{\rm i}\gg{}l_\perp$ and $R_{\rm o}\rightarrow{}\infty$ we have ${\cal A}\rightarrow{}3/2$, and such a~limit can be regarded as an additional conductance minimum. The asymptotic form of Eq.\ (\ref{gblg0dsk}) then reads
\begin{equation}
  \label{gblgasym}
  \frac{G_{\rm BLG}(t'=0)}{g_0} \simeq \frac{2R_{\rm i}^3}{l_\perp^2 R_{\rm o}}
  \ \ \ \ (R_{\rm i}\gg{}l_\perp,R_{\rm o}\rightarrow{}\infty).
\end{equation}
Although qualitative features of the quantum-tunneling regime for the Corbino geometry are reproduced [in particular, $G_{\rm BLG}(t'=0)\propto{}R_{\rm o}^{-1}$], the asymptotic conductance is elevated by a~large factor of $R_{\rm i}^2/(4\l_\perp^2)$ in comparison to the MLG disk case \cite{gasymfoo}.

\begin{figure}
\includegraphics[width=0.9\linewidth]{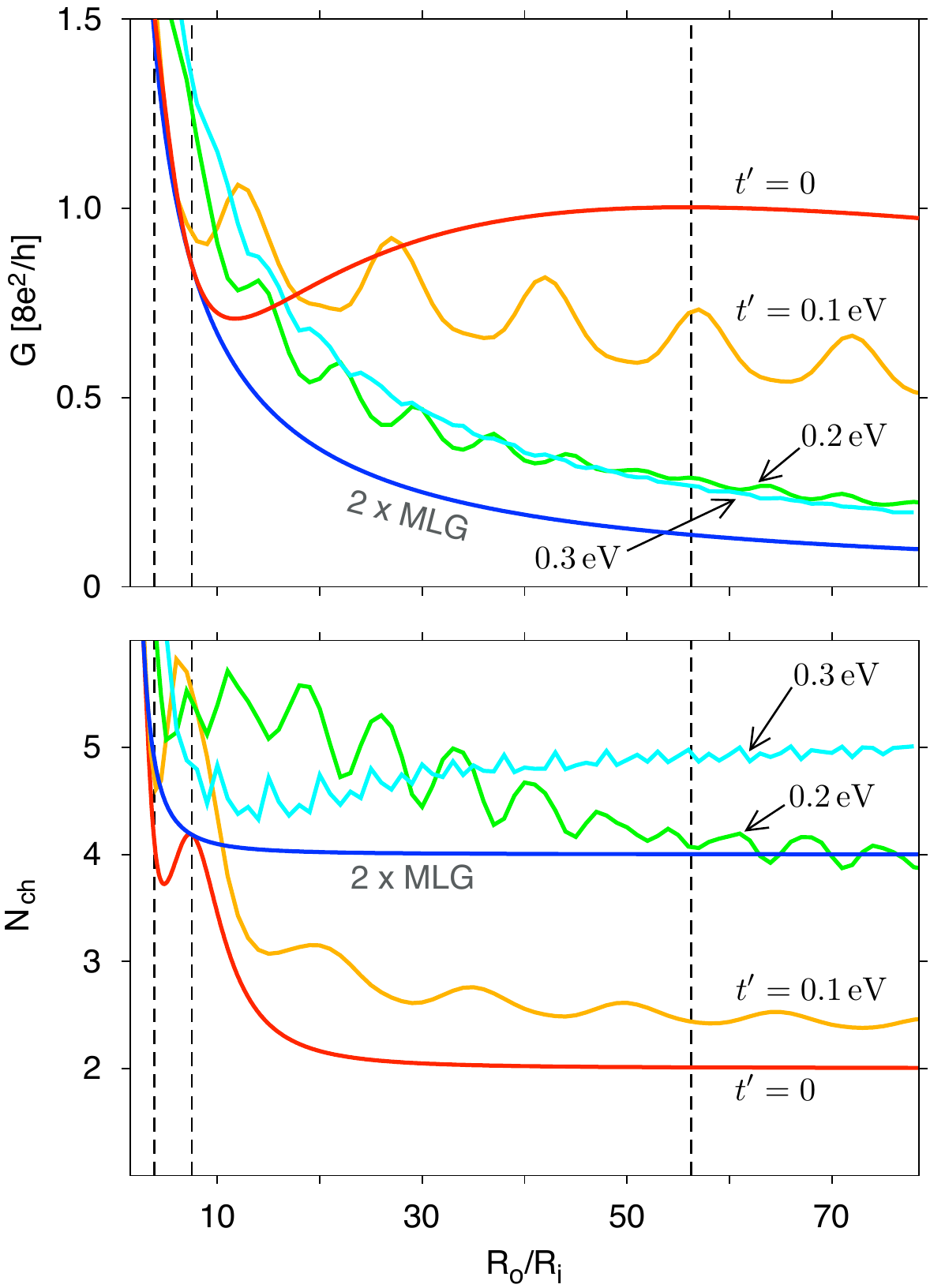}
\caption{ \label{gblgzerofi}
Radii-ratio dependence of the conductance $G=2\pi{}\sigma/\ln(R_{\rm o}/R_{\rm i})$ (top panel) and the effective number of transmission channels (bottom panel) at a~fixed $R_{\rm i}=15\,\l_\perp\approx{}24\,$nm and varying $t'$ (specified for each line). Vertical lines mark the values of ${\cal A}\approx{}3$ (left), ${\cal A}\approx{}5/2$ (middle), and ${\cal A}\approx{}2$ (right) following from Eq.\ (\ref{calaappr}). The corresponding results for two decoupled MLG disks ($t_\perp=t'=0$) are also shown. 
}
\end{figure}

In Fig.\ \ref{gblgzerofi}, we compare the conductance (in the top panel) and the effective number of transmission channels  (in the bottom panel)
\begin{equation}
  \label{nchandef}
  N_{\rm ch}=\frac{G}{g_0\left(1-{\cal F}\right)}=
  \frac{\left(\sum_l T_l\right)^2}{\sum_l T_l^2},
\end{equation}
with the index $l\equiv(m,\pm)$ accounting for angular-momenta and layer degrees of freedom, following from Eqs.\ (\ref{gblg0dsk}--\ref{upsidef}) [red lines] for $t'=0$; as well as for the case of decoupled MLG disks ($t_\perp=t'=0$) [blue lines]; with the numerical results obtained from Eqs.\ (\ref{landsig}) and (\ref{landff}) for $t'=0.1\,$eV, $0.2\,$eV, and $0.3\,$eV [remaining lines]. The inner radius is fixed at $R_{\rm i}=15\,l_\perp$, and the outer radius is varied in the range of $1.5<R_{\rm o}/R_{\rm i}<80$. 
The effects of trigonal warping are visible for all radii ratios considered, and become particularly significant when approaching $R_{\rm o}/R_{\rm i}=R_{\rm i}^2/(4l_\perp^2)\approx{}56$, corresponding to ${\cal A}\approx{}2$ [following from Eq.\ (\ref{calaappr})]. In such a~parameter range, $G$ systematically decreases, whereas $N_{\rm ch}$ systematically increases when enlarging $t'$. We also notice that Fabry-Perot resonances, corresponding to integer $k_\ell(R_{\rm o}-R_{\rm i})$, are visible for $t'\neq{}0$, indicating the contribution from secondary Dirac points. 

For smaller radii ratios, including $R_{\rm o}/R_{\rm i}=7.5$ (corresponding to ${\cal A}\approx{}5/2$) and  $R_{\rm o}/R_{\rm i}=3.83$ (corresponding to ${\cal A}\approx{}3$) [see vertical lines in Fig.\ \ref{gblgzerofi}], the system is close to the pseudodiffusive charge-transport regime. In the $t'=0$ case, the conductance minimum  is shifted from $R_{\rm o}/R_{\rm i}=7.5$ to $R_{\rm o}/R_{\rm i}\approx{}10$ due to the influence of transmission channels with higher angular momenta. (We further notice that the effective number of channels has a~local maximum $N_{\rm ch}\approx{}4.2$ at $R_{\rm o}/R_{\rm i}\approx{}7.5$, where it also precisely matches the value for two decoupled MLG disks, in good agreement with predictions for a~quantum-tunneling regime reported earlier in this subsection.) The trigonal warping noticeably enhances the conductance for $R_{\rm o}/R_{\rm i}\lesssim{}10$; i.e., the effect is opposite then for larger $R_{\rm o}/R_{\rm i}$; with some exception for the smallest considered $t'=0.1\,$eV and $R_{\rm o}/R_{\rm i}\lesssim{}7$, as the disk diameter $2R_{\rm o}\lesssim{}\ell$ in such a~case. 

The evolution of $G$ and $N_{\rm ch}$ with the trigonal warping strength, illustrated in Fig.\ \ref{gblgzerofi}, clearly shows that the role of two transmission channels with angular momenta $\pm{}2\hbar$, prominent for $t'=0$ and large $R_{\rm o}/R_{\rm i}$, is strongly suppressed for $t'\neq{}0$, indicating the gradual crossover to a~quantum tunneling regime characterized by $G\propto{}R_{\rm o}^{-1}$. We attribute it to the fact that a~$d$-wave symmetry of normal modes in leads with $m=\pm{}2$ does not match the $D_{3d}$ dihedral symmetry of the low-energy Hamiltonian for $t'\neq{}0$. Moreover, for the two largest considered values of $t'=0.2\,$eV and $0.3\,$eV, both $G$ and $N_{\rm ch}$ are noticeably amplified in comparison to the relevant characteristics for two decoupled MLG disks, signalling the role of quantum states close to the secondary Dirac points is important when the crossover to a~quantum-tunneling regime occurs.

\section{Magnetotransport characteristics}

\subsection{Analytically soluble disk-shaped systems}
Our discussion of the magnetotransport characteristics starts from pointing out that the influence of a~uniform magnetic field $B$  \cite{unifoo} on analytical results given by Eqs.\ (\ref{gmlgdisk}) and (\ref{gblg0dsk}) can be expressed in a~very compact form: Namely, it is sufficient to replace $j$ and $m$ in arguments of hyperbolic cosine by \cite{Kat10,Ryc10,Rut14d}
\begin{equation}
  \overline{\jmath} = j + \frac{\Phi_D}{\Phi_0} \ \ \ \text{and} \ \ \ 
  \overline{m} = m + \frac{\Phi_D}{\Phi_0},
\end{equation}
with 
\begin{align}
  \Phi_D &= \pi{}B(R_{\rm o}^2-R_{\rm i}^2) \label{phddef} \\
  \text{and} \ \ \ \ 
  \Phi_0 &= \frac{2h}{e}\ln(R_{\rm o}/R_{\rm i}), \label{ph0def}
\end{align}
denoting the flux piercing the disk area ($\Phi_D$) and the basic period of magnetoconductance oscillations ($\Phi_0$). 

All the considered charge-transfer characteristics $\sigma$ (\ref{landsig}), ${\cal F}$ (\ref{landff}), and ${\cal R}$ (\ref{landrr}), are predicted theoretically to show periodic oscillations with the magnetic flux $\Phi_D$ [see Eq.\ (\ref{phddef})] piercing a~graphene-based Corbino disk \cite{Ryc10,Kat10,Rut14d,Rut15a}, provided the Fermi energy corresponds to the Dirac point or to any other Landau level (LL). The oscillations appear due to a~limited number of transmission channels for $R_{\rm o}\gg{}R_{\rm i}$, and show a~formal analogy with similar effects discussed for a nanotube in a magnetic field applied along the axis \cite{Kol12}. One should notice, however, that the oscillation period $\Phi_0$ [see Eq.\ (\ref{ph0def})] for a~disk in a~uniform, perpendicular field, corresponds to a~physical field of $18\,$mT for a~typical $1\,\mu$m disk ($R_{\rm o}=5R_{\rm i}=500\,$nm), while for a~$1\,$nm diameter nanotube in axial field the period, given by the standard Aharonov-Bohm flux quantum $\Phi_{\rm AB}=h/e$, corresponds to $B\approx{}5300\,$T. 

Similar effects  were also considered for graphene disks with strain-induced pseudomagnetic fields \cite{Kha13} and with the spin-orbit coupling \cite{Vil14}, extending the list of different theoretical proposals for producing a valley polarization with graphene-based nanostructures \cite{Rec07,Ryc07,Akh08,Fuj10,Gun11,Gru14} and related systems \cite{Pal11,Cul12,Gon13,HXu15}.

For MLG disks, the oscillations magnitude depends only on the radii ratio $R_{\rm o}/R_{\rm i}$. For the conductivity, we have $\Delta{}\sigma_{\rm MLG}\gtrsim{}0.1\sigma_{\rm MLG}$ if $R_{\rm o}/R_{\rm i}\geqslant{}5$. For BLG disks, the mode-matching analysis for the $t'=0$ case \cite{Rut14d} unveils an interference between the two transmission channels for each angular momentum eigenvalue, following from the coupling between the layers quantified by $t_\perp$. In turn, the oscillations magnitude depends also on the physical system size. For instance, conductivity oscillations are predicted to vanish ($\Delta\sigma=0$) for
\begin{equation}
  \label{condvani}
  \frac{R_{\rm o}}{R_{\rm i}}\simeq{}
  \left(\frac{R_{\rm i}t_\perp}{2\hbar{}v_F}\right)^{4/p}
  \ \ \ \ (\text{for }R_{\rm o}\gg{}R_{\rm i}),
\end{equation}
where $p=1,2,3,\dots$. Eq.\ (\ref{condvani}) with {\em even} $p$ [equivalent to half-odd integer $q$ in Eq.\ (\ref{maxgblg})] gives the condition for maximal oscillations, with the magnitude $\Delta\sigma=2\Delta\sigma_{\rm MLG}$, same in the limit of decoupled layers ($t_\perp\rightarrow{}0$).

If the Fermi energy $E$ is close but not precisely adjusted to the Dirac point, the oscillations in both MLG and BLG disks are still predicted to appear in the limited range of magnetic fluxes, namely
\begin{equation}
  \label{phidmax}
  |\Phi_D|\leqslant{}\Phi_D^{\rm max}\simeq{}-\frac{2h}{e}\ln(k_FR_{\rm i}), 
  \ \ \ \ (\text{for }k_FR_{\rm i}\ll{}1),
\end{equation}
with $k_F=|E|/(\hbar{}v_F)$, away from which the conductivity is strongly suppressed. Eq.\ (\ref{phidmax}) can be rewritten to obtain the corresponding energy range for a given field
\begin{equation}
  \label{efermax}
  |E|\lesssim{} \frac{\hbar{}v_F}{R_{\rm i}}
  \exp\left( -\frac{R_{\rm o}^2-R_{\rm i}^2}{l_B^2} \right). 
\end{equation}
The limits given by Eqs.\ (\ref{phidmax}) and (\ref{efermax}) essentially applies to higher charge-transfer cumulants as well, albeit the dimensionless characteristics ${\cal F}$ and ${\cal R}$ were recently found to show stable, quasiperiodic oscillations in the high source-drain voltage limit \cite{Rut15a}.

Later this Section, we employ the numerical procedure described in Section II in order to find out how the magnetotransport characteristics of BLG disks are affected by the trigonal warping ($t'\neq{}0$).

\subsection{Rectangular BLG device with periodic boundary conditions}
Before discussing magnetotransport of the Corbino disk, it is instructive to consider a simpler artificial system depicted schematically in Fig.\ \ref{setcoraux}(b). A BLG strip of width $W$, contacted by the electrodes at a distance $L$, in a~uniform field $B$, and with periodic boundary conditions in the transverse direction, was earlier discussed in the $W\gg{}L$ limit \cite{Rut14a}, in which the pseudodiffusive charge transport is predicted to appear near LLs. Here we primarily focus on the $W\lesssim{}\pi{}L$ range (a nanotube-like geometry) which do not seem to have a~direct physical analogue, but can be treated in analytical terms and possesses a discrete spectrum of transmission channels closely resembling the situation in the Corbino disk. 

The wavefunctions for a rectangular sample are presented in Appendix~B. Each spinor component can be written as a product of the exponential function and the Airy function, with their arguments scaling as $l_B^{-2}\propto{}B$ for high fields; see Eq.\ (\ref{phiksamp}). In turn, taking the asymptotic form of the Airy function $\mbox{Ai}\left(z\right)\simeq\mbox{exp}\left(-2z^{3/2}/3\right)/\left(2z^{1/4}\sqrt{\pi}\right)$, we find the conductivity in the high-field limit can roughly be approximated by
\begin{multline}
  \sigma_{L\gg{}l_B} \approx \frac{g_0L}{W}\sum_{k,\pm}
  \mbox{cosh}^{-2}\left\{ L\left(k-\frac{L}{2l_{B}^{2}}\right)\right. \\
  \left.\pm\frac{1}{24}\Re\left[\gamma_{k}\left(L\right)
      -\gamma_{k}\left(0\right)\right]\right\}, \label{eq:Gapprox} 
\end{multline}
where $k=0,\pm{}2\pi/W,\pm{}4\pi/W,\dots$, 
\begin{equation}
  \label{eq:gammak}
  \gamma_{k}\left(x\right)=
  \sqrt{-\frac{i\nu{}l_B}{l_\perp}}\, \left(\frac{8x}{l_B}-8kl_B-\frac{i\nu{}l_B}{l_\perp}\right)^{3/2},
\end{equation}
and $\Re\left(z\right)$ denotes the real part of $z$. In the absence of trigonal warping, Eq.\ (\ref{eq:Gapprox}) is replaced by exact expression for the conductivity, namely
\begin{multline}
  \label{eq:Gexact}
  \frac{\sigma_{\rm rec}(t'\!=\!0)}{\sigma_0} =
  1 + \sum_{n=1}^{\infty}{\Bigg\{} \left(\frac{\pi{}nW}{L}\right)\cos(nk_cW) \\
  \times\left.
  \left[{\sinh\left(\displaystyle\frac{\pi{}nW}{2L}\right)}\right]^{-1}
  \cos\left(\frac{\pi{}ne}{h}LWB\right)\right\},
\end{multline}
where $k_c=(1/L)\,\mbox{arsinh}\left[\,Lt_\perp/(2\hbar{}v_F)\,\right]$. 
A~Fourier series on the right-hand side represents periodic (approximately sinusoidal for moderate aspect ratios $W/L\gtrsim{}2$) magnetoconductance oscillations with the period ${\cal T}_0=(2h/e)(LW)^{-1}$ \cite{Jackiw}. 

For $t'\neq{}0$, the approximating expression given by Eqs.\ (\ref{eq:Gapprox},\ref{eq:gammak}) is particularly convenient when extracting the beating frequency which governs high-field magnetotransport characteristics of the system. It can be shown that the presence of two relevant transmission eigenvalues for each momentum $k$ in Eq.\ (\ref{eq:Gapprox}) follows directly from the fact that the lowest LL in BLG has an additional twofold degeneracy absent for higher LLs \cite{Rut14a}. We define the two field-dependent effective system sizes determining transmission probabilities in Eq. (\ref{eq:Gapprox}), namely
\begin{equation}
  \label{lenpmdef}
  L_\pm=L\mu_\pm,
\end{equation}
and the corresponding momentum-quantization shifts
\begin{equation}
  \label{delkpm}
  \Delta{}k_\pm=\frac{L}{2l_B^2\mu_\pm},
\end{equation}
where
\begin{equation}
  \mu_\pm=1\pm{}
  \sqrt{\frac{8\nu l_B^2}{9Ll_\perp}}.
\end{equation}

The above reasoning holds true for high magnetic fields and arbitrary $W/L$ ratio. For instance, the conductivity in the $W\gg{}L$ limit can be approximated by
\begin{equation}
  \label{eq:Gapprox2}
  \sigma_{W\gg{}L\gg{}l_B}\approx{}
  \frac{2g_0}{\pi{}}\left(\frac{1}{\mu_+}+\frac{1}{\mu_-}\right)=
  \sigma_0 \left(1-\frac{8\nu l_B^2}{9Ll_\perp}\right)^{-1}, 
\end{equation}
restoring the $t'=0$ value ($\sigma\rightarrow\sigma_0$) for $l_B\rightarrow{}0$.

\begin{figure}
\includegraphics[width=0.8\linewidth]{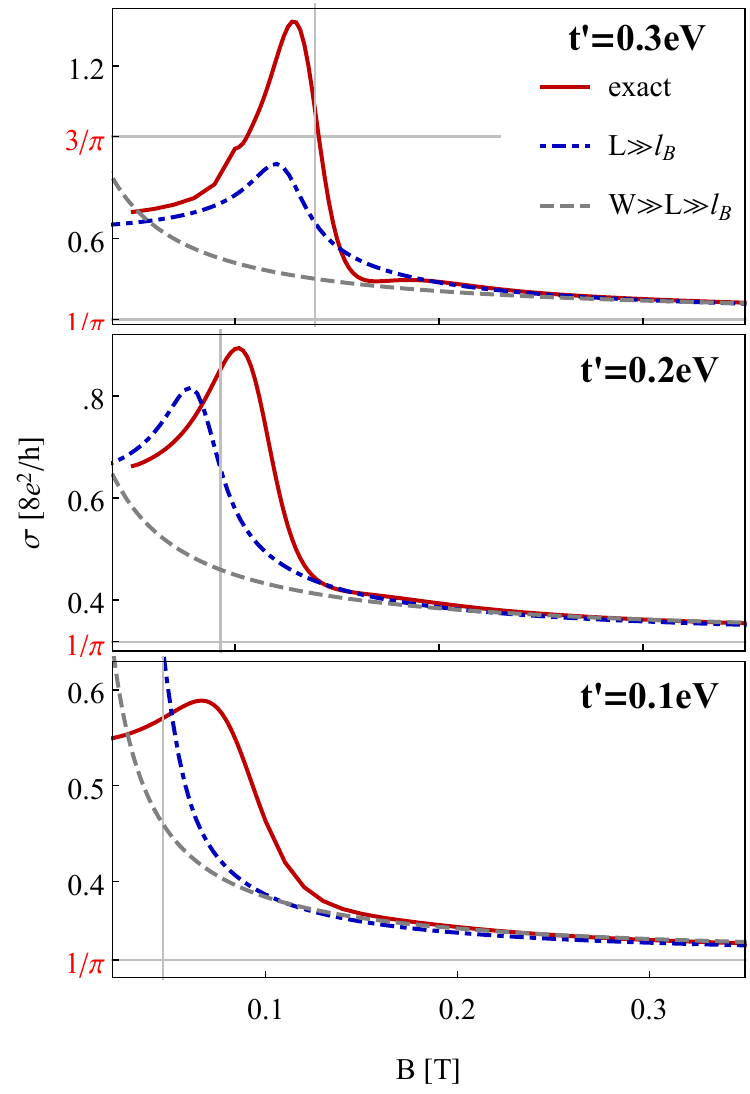}
\caption{\label{reclowfi}
  Conductivity of the rectangular BLG sample at the charge-neutrality point as a function of the magnetic field $B$. The sample length is fixed at $L=300\,l_\perp\simeq{}480\,$nm, the width is $W=10L$, and the skew-interlayer hopping integral $t'$ is varied between the panels. {Red solid lines:} exact numerical results obtained from the mode-matching analysis via Eq.\ (\ref{landsig}). {Blue solid lines:} results from  Eqs.\ (\ref{eq:Gapprox},\ref{eq:gammak}) for $L\gg{}l_B$. {Dashed lines:} results from Eq.\ (\ref{eq:Gapprox2}) for $W\gg{}L\gg{}l_B$. 
}
\end{figure}

In Fig.\ \ref{reclowfi} we compare exact numerical results obtained from Eq.\ (\ref{landsig}) for $W=10L=3000\,l_\perp$ and different values of skew-interlayer hopping $t'$ [red solid lines] with corresponding results following from the approximating Eqs.\ (\ref{eq:Gapprox},\ref{eq:gammak}) [blue solid lines] and Eq.\ (\ref{eq:Gapprox2}) [dashed lines]. For low magnetic fields, the conductivity monotonically grows with increasing $B$ (for any $t'\neq{}0$), up to the maximal value (at $B=B_{\rm peak}$) which may exceed $3\sigma_0$ for larger $t'$-s. For $B>B_{\rm peak}$, the effect of trigonal warping on the conductivity is gradually suppressed, leading to $\sigma-\sigma_0\propto{}t'/B$, in agreement with the approximating Eq.\ (\ref{eq:Gapprox2}). 

The value of $B_{\rm peak}$ is related to the quasimomentum shifts given by Eq.\ (\ref{delkpm}). For a finite aspect ratio $W/L$, maximal conductivity appears when the average shift is of the same order of magnitude as the distance between primary and secondary Dirac points [given by Eq.\ (\ref{elldef})], namely
\begin{equation}
  \left(\frac{\Delta{}k_++\Delta{}k_-}{2}\right)_{\rm peak}\sim{}\,k_\ell,
\end{equation}
what can be rewritten as
\begin{equation}
  \label{bpeakrect}
  B_{\rm peak}\sim{}\frac{2h}{e}\frac{1}{L{\ell}}\equiv{}B_L,
\end{equation}
where  we have defined the crossover {\em Lifshitz field} $B_L$. It is also visible in the top panel of Fig.\ \ref{reclowfi} that the approximating Eqs.\ (\ref{eq:Gapprox},\ref{eq:gammak})  [blue dashed lines] reproduces the peak position (albeit not the maximal conductivity) with a~good accuracy for $t'=0.3\,$eV. 

The conductivity maximum at $B_{\rm peak}\sim{}B_L\neq{}0$ can be regarded as the first effect of the trigonal warping, appearing for samples with finite aspect ratios, but still well-visible for $W/L=10$.

\begin{figure}
\includegraphics[width=0.9\linewidth]{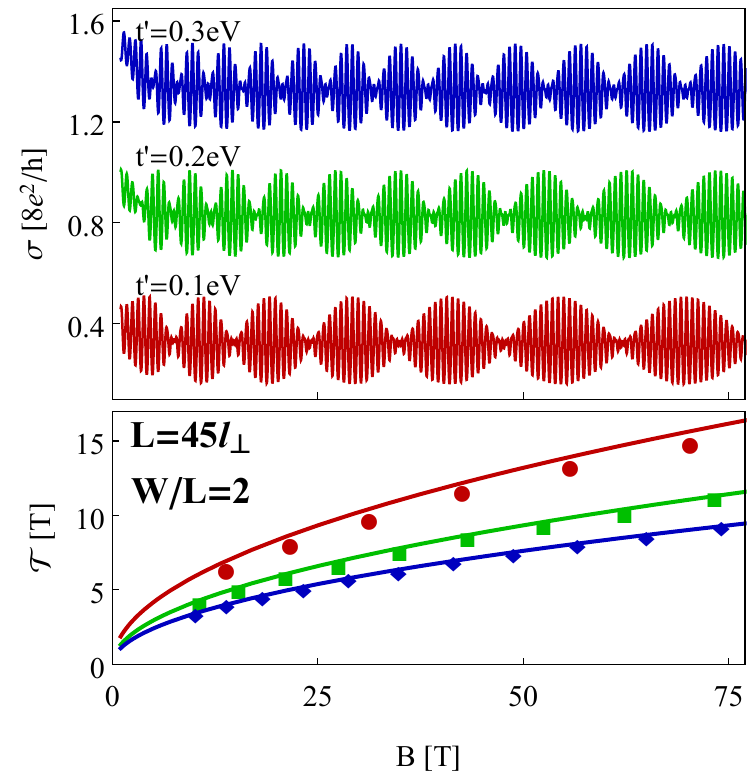}
\caption{\label{rectbeats}
  Magnetoconductance oscillations for the rectangular BLG sample of the length 
$L=45\,l_\perp\simeq{}72\,$nm and the width $W=2L$ at the charge-neutrality 
point. Top panel: The conductivity as a function of magnetic field for three 
values of the skew-interlayer hopping integral $t'$ (specified for each line). 
The lines for $t'=0.3\,$eV and $t'=0.2\,$eV are shifted by $1$ and $1/2$ 
for clarity.
Bottom panel: 
Consecutive periods of the beating envelope extracted from the data
shown in the top panel, for $t'=0.1\,$eV (circles), $t'=0.2\,$eV (squares), 
and $t'=0.3\,$eV (diamonds).
Solid lines correspond to the approximating Eq.\ (\ref{eq:envper}).
}
\end{figure}

The second effect, present in systems with $W\lesssim{}\pi{}L$, is the emergence of beating patterns. In the top panel of Fig.\ \ref{rectbeats} we display the conductivity, as a~function of magnetic field, for $W=2L=90\,l_\perp$. Quasiperiodic beatings are characterized by the field-dependent envelope period, which can be approximated by
\begin{equation}
  \label{eq:envper}
  \mathcal{T}(B) \approx \frac{4h}{eLW}\frac{\mu_{+}\mu_{-}}{\mu_{+}-\mu_{-}}
  \simeq  \frac{3h}{e}\frac{1}{LW}\left(\frac{B}{B_L}\right)^{1/2},
\end{equation}
while the period of internal oscillations remains the same as in Eq.\ (\ref{eq:Gexact}).  The comparison between the actual envelope periods refined from the numerical data [datapoints] and the values following from Eq.\ (\ref{eq:envper}) [lines] is provided in the bottom panel of Fig.\ \ref{rectbeats}. 

For high magnetic fields, Eq.\ (\ref{eq:envper}) leads to $\mathcal{T}(B)\propto\sqrt{B}$. In turn, the $t'=0$ behavior characterized by single-frequency, sinusoidal magnetoconductance oscillations with a~size-dependent amplitude [see Eq.\ (\ref{eq:Gexact})] cannot be restored for any $t'\neq{}0$ in the high-field limit. We interpret this effect as a~finite-system analog of the zero-temperature conductance instability (the parameter-driven Lifshitz transition)  in bulk BLG samples \cite{Cse07a}.

\begin{figure}
  \includegraphics[width=0.8\linewidth]{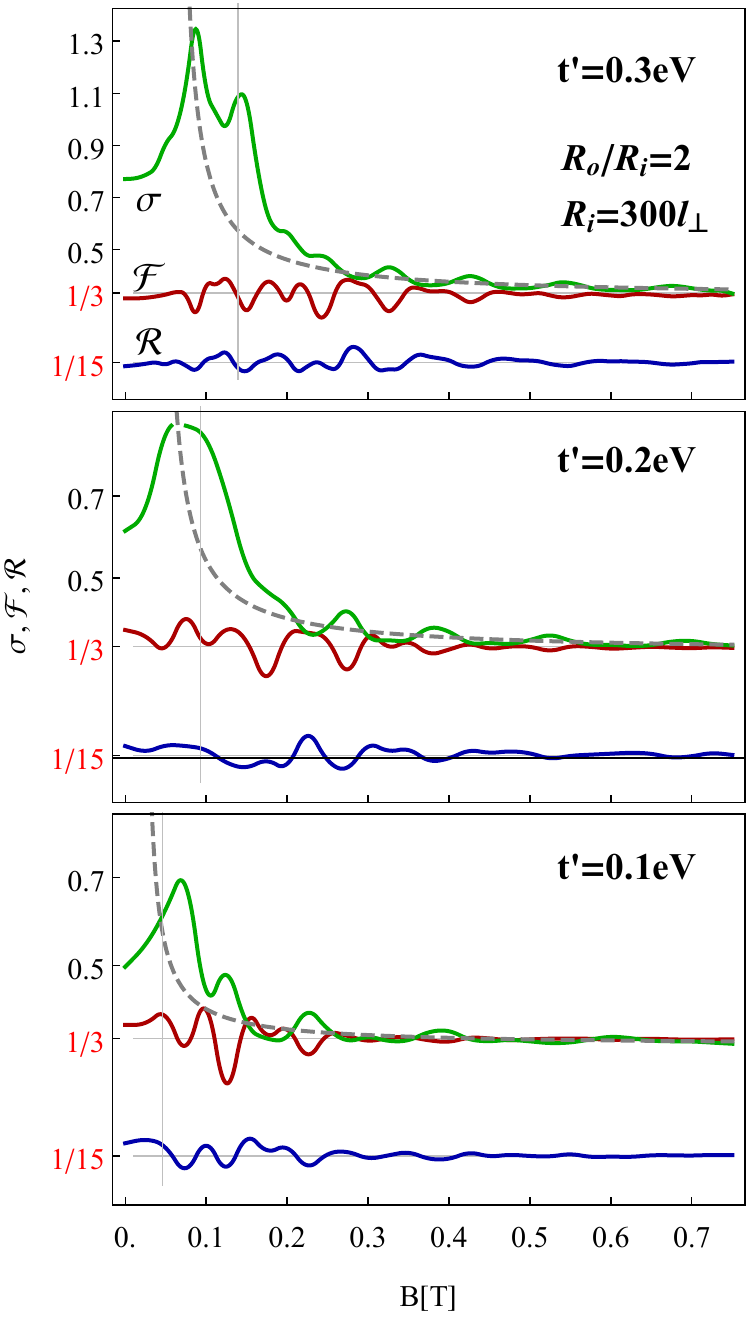}
  \caption{\label{pdiffdisk}
    Conductivity $\sigma$ [$\,$specified in the units of $8e^2/h\,$], the Fano 
    factor $\mathcal{F}$, and the $\mathcal{R}$ factor for BLG Corbino disk 
    with $R_{\rm i}=300\,l_\perp$, $R_{\rm o}=2R_{\rm i}$, and different 
    skew-interlayer hopping integrals $t'$ (specified for each panel). 
    Vertical lines correspond to $B=B_L$ [see Eq.\ (\ref{bpeakcorb})], 
    the horizontal lines mark the pseudodiffusive values
    ${\cal F}=1/3$ and ${\cal R}=1/15$, the dashed lines
    represent the approximating Eq. (\ref{eq:przewBC}). 
  }
\end{figure}

\begin{figure}[h]
  \includegraphics[width=0.9\linewidth]{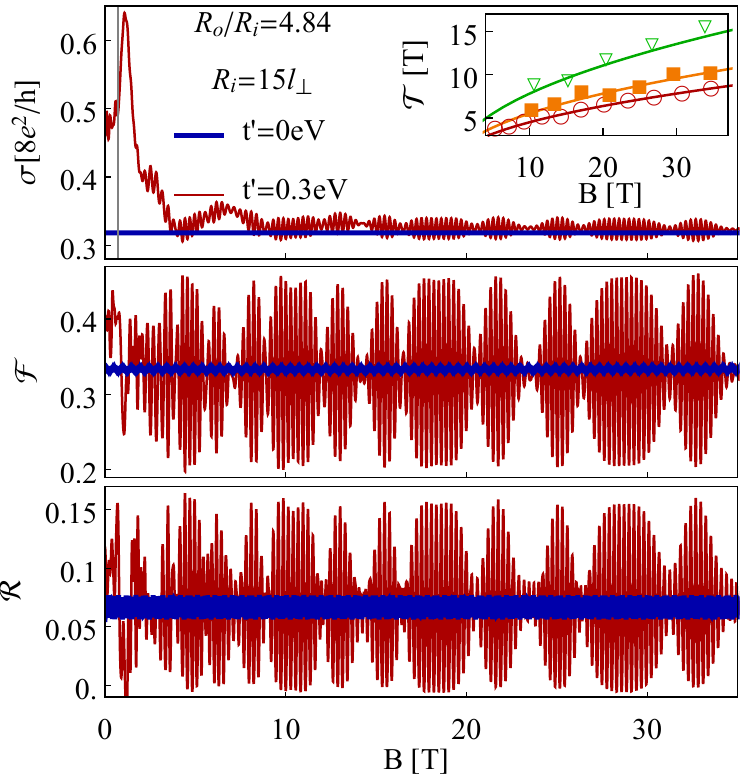}
  \caption{\label{qbeatdisk}
    Charge-transfer characteristics $\sigma$, $\mathcal{F}$, and $\mathcal{R}$ 
    for BLG disk with $R_{\rm i}=15\,l_\perp$, $R_{\rm o}=4.84\,R_{\rm i}$. 
    Each of the main panels compares the numerical results for $t'=0$ 
    (blue line) and $t'=0.3\,$eV (red line). 
    The vertical grey line in the top panel marks the value of $B_L$ obtained 
    from Eq.\ (\ref{bpeakcorb}) for $t'=0.3\,$eV. 
    The inset: Period of the beating envelope for three different
    values of $t'=0.3\,$eV (circles), $0.2\,$eV
    (orange squares), $0.1\,$eV (triangles). 
    Solid lines correspond to Eq. (\ref{eq:okrCor}).
  }
\end{figure}

\subsection{BLG Corbino disks}
Our numerical study of magnetotransport through BLG Corbino disks focuses on two different systems: First one, characterized by $R_{\rm i}=300\,l_\perp$ and $R_{\rm o}=2R_{\rm i}$ is in the pseudodiffusive charge-transport regime, whereas the second one, with $R_{\rm i}=15\,l_\perp$ and $R_{\rm o}=4.84\,R_{\rm i}$, shows the beating patterns. In the latter case, the parameters are chosen such that the magnetoconductance oscillations vanish for $t'=0$ (see Ref.\ \cite{Rut14d}), in order to illustrate the role of trigonal warping more clearly. 

The numerical results for $\sigma$, ${\cal F}$, and ${\cal R}$ are presented, as functions of magnetic field $B$, in Figs.\ \ref{pdiffdisk} and \ref{qbeatdisk}. 

In the pseudodiffusive transport regime (see Fig.\ \ref{pdiffdisk}) some irregular fluctuations, visible for all the discussed charge-transfer characteristics (and all three values of $t'$), are suppressed for magnetic fields $B\gtrsim{}0.5\,$T. We attribute these fluctuations rather to a~Fabry-Perot interference than to the angular momentum quantization. A striking feature of the data presented in Fig.\ \ref{pdiffdisk} is that a distinct conductance peak appears at each panel near the field 
\begin{equation}
  \label{bpeakcorb}
  B_L= \frac{4}{3}\sqrt{3}\,
  \frac{\hbar{}\,t't_\perp}{e\,t_0^2a\left(R_{\rm o}-R_{\rm i}\right)},
\end{equation}
closely resembling the phenomena described above for a~rectangular sample. [Notice that the rightmost equality in Eq.\ (\ref{bpeakrect}) and Eq.\ (\ref{bpeakcorb}) are equivalent provided that $L\equiv{}R_{\rm o}-R_{\rm i}$]. Above the crossover field $B_L$, the conductivity can be approximated by the formula
\begin{equation}
  \label{eq:przewBC}
  \sigma_{B\gg{}B_L}\approx{}
  \sigma_0\left(1-0.5\,\frac{B_L}{B}\right)^{-1},
\end{equation}
which is visualized with  dashed lines in Fig.\ \ref{pdiffdisk},
and can be regarded as a version of Eq.\ (\ref{eq:Gapprox2}) for BLG disk. 

For the disk with larger radii ratio (see Fig.\ \ref{qbeatdisk}) all three magnetotransport characteristics exhibit quasiperiodic beating patterns for $t'=0.3\,$eV. This is not the case for $t'=0$, when each characteristic shows approximately sinusoidal, single-frequency, oscillations with a~constant amplitude. As the system no longer possesses a~rotational symmetry in the presence of trigonal warping, the magnetotransport cannot be simply rationalized by defining two effective system sizes in analogy to Eq.\ (\ref{lenpmdef}). Comparing to the rectangular system case (which can be regarded, due to the periodic boundary conditions, as a~nanotube-like, i.e., possessing the rotational symmetry also for $t'\neq{}0$) illustrated in Fig.\ \ref{rectbeats}, the beatings are slightly less regular now, clustering in the groups of three. This certain feature of the data displayed in Fig.\ \ref{qbeatdisk} suggests the presence of four, rather than two, quasiperiodic components determining the transmission probabilities. 

It is worth to stress here that the main features of the magnetotransport characteristics still resemble the rectangular (or a~nanotube-like) system case. In particular, we found that the beating-envelope period can now be approximated by
\begin{equation}
  \label{eq:okrCor}
  {\cal T}(B)\approx
  \frac{6.9}{\pi(R_{\rm o}^2-R_{\rm i}^2)}\,\frac{h}{e}
  \left(\frac{B}{B_L}\right)^{1/2},
\end{equation}
see the inset in Fig. \ref{qbeatdisk}. On the other hand, the period of the internal oscillations, ${\cal T}_0=\Phi_0/\left[\pi({R_{\rm o}}^2-{R_{\rm i}}^2)\right]$, remains the same as the basic period for MLG disks or BLG disks with $t'=0$ \cite{Ryc10,Kat10,Rut14d}. Also, a high-field behavior of the conductivity averaged over consecutive intervals, each one of the ${\cal T}_0$ width, can be approximated by Eq.\ (\ref{eq:przewBC}).


\section{Conclusions}
We have investigated, by means of numerical transfer-matrix approach in the angular-momentum space, the effects of the skew-interlayer hopping integrals (the trigonal warping) on selected transport characteristics of bilayer-graphene (BLG) Corbino disks. Additionally, the analytical mode-matching  for an artificial (nanotube-like) system, formed of a~BLG strip upon applying the periodic boundary conditions, was briefly presented and the analogies between these two systems were put forward. 

If the Fermi energy is close to the charge-neutrality point, both the scaling behavior at zero magnetic field (which would require a comparison between devices of different sizes in an experimental study) and the single-device magnetotransport discussion unveils several phenomena, in which transport characteristics, such as the conductivity, the Fano factor, and the third charge-transfer cumulant, are noticeably affected by the trigonal warping.

In the pseudodiffusive transport regime, corresponding to the disk radii ratios $R_{\rm o}/R_{\rm i}\lesssim{}2$, the conductivity shows a~one-parameter scaling, in agreement with predictions of Ref.\ \cite{Rut14b} for a~rectangular sample. In the Corbino geometry, however, the role a~crystallographic orientation is eliminated, and the zero-field minimal conductivity can be approximated by
\begin{equation}
  \sigma_{\rm min}\approx{}3\sigma_0\left[\,1-\left(
      \frac{\lambda}{R_{\rm o}\!-\!R_{\rm i}}\right)^{0.5}\,\right],
\end{equation}
where $\lambda=\lambda(t')$ depends only on the skew-interlayer hopping $t'$, and varies from $\lambda\approx{}20\,$nm for $t'=0.3\,$eV to $\lambda\approx{}100\,$nm for $t'=0.1\,$eV. In the uniform magnetic field $B$, the conductivity increases reaching the maximal value $\sigma_{\rm max}\gtrsim{}3\sigma_0$ near the so-called Lifshitz field $B_L$, for which the magnetic length follows the relation
\begin{equation}
  \left(R_{\rm o}\!-\!R_{\rm i}\right)\ell = 4\pi{}l_B^2,
\end{equation}
where $\ell=\sqrt{3}\,\pi{}at_0^2/(t_\perp{}t')$ is defined by $t'$ and other microscopic parameters: the lattice spacing $a$ as well as the nearest-neighbor intra- and interlayer hoppings $t_0$ and $t_\perp$. 
Above $B_L$, the conductivity gradually decreases showing the asymptotic behavior $\sigma-\sigma_0\propto{}B_L/B$ [see Eq.\ (\ref{eq:przewBC})]. The second and third charge transfer cumulants stay close to their pseudodiffusive values (${\cal F}=1/3$, ${\cal R}=1/15$) when varying the system size or the magnetic field. 

In the opposite, quantum-tunneling regime (corresponding to $R_{\rm o}/R_{\rm i}\gtrsim{}4$), the charge-transfer characteristics are also sensitive to $t'$. At zero field and $t'=0$, the transport is governed by two quantum channels with angular momenta $\pm{}2\hbar$. For $t'\neq{}0$, the backscattering is enhanced for these channels, as the related wavefunctions no longer match the symmetry of the low-energy Hamiltonian. At high magnetic fields, all the charge-transfer characteristics show quasiperiodic beating patterns, with the envelope period ${\cal T}(B)\propto{}\sqrt{B/B_L}$ [see Eq.\ (\ref{eq:okrCor})]. Most remarkably, the beating patterns, triggered by the trigonal-warping ($t'\neq{}0$), remain well-pronounced in the Landau quantization regime. (Unlike the average conductivity enhancement, which is usually eliminated by a~few Tesla field.) It seems this finite-system version of the Lifshitz transition can be related to numerous phenomena appearing in different branches of physics, starting from semiconducting heterostructures \cite{Gui04}, via strongly-correlated electron systems \cite{Kor95}, to neutrino physics \cite{Pon67}, in which scattering the particles between quantum states with different effective masses leads to oscillations in relevant counting statistics, although this time the interference occurs between the evanescent waves. 

We stress here that finding the Lifshitz field $B_L$, via the asymptotic behavior of the conductivity, or via the beating period, may allow one -- at least in principle -- to determine the value of $t'$ from a single-device magnetotransport measurement. 

Apart from the possible verification of tight-binding parameters in BLG Hamiltonian, we believe the effects we describe, when confirmed experimentally, will provide a~thorough insight into the interplay between massless- and massive-chiral states ruling the quantum transport through BLG devices near the charge-neutrality point. For instance, the conductivity enhancement for $B\sim{}B_L$ may dominate the signatures of interaction-related magnetic catalysis phenomenon \cite{Gus94} (particularly in finite-size systems), and one should precisely distinguish single- and many-body aspects when searching for this intriguing phenomenon in BLG. 

As we have focused on clean ballistic systems, several factors which may modify the transport properties of graphene-based devices, including the disorder \cite{Mac13}, lattice defects \cite{Dre10}, or magnetic impurities \cite{Cor09,Sza11,Hon13}, are beyond the scope of this work. Some experimental \cite{Mia07,May11} and numerical \cite{Ort13,Gat14} findings suggest that charge-transfer characteristics in the pseudodiffusive transport regime are quite robust against such factors. For the opposite, quantum-tunneling regime, we put forward the following reasoning: In the presence of trigonal warping, the rotational symmetry supposed in earlier studies of MLG \cite{Ryc10,Kat10} or BLG \cite{Rut14d} disks, no longer applies. In spite of this fact, the basic oscillation period [$\,\Phi_0=2\,(h/e)\ln(R_{\rm o}/R_{\rm i})$, in terms of magnetic flux piercing the disk area] remains unaltered, allowing one to believe that oscillations and beating patterns would appear in a more general situation as well.

\section*{Acknowledgements}
We thank Stephan Roche for the correspondence. 
The work was supported by the National Science Centre of Poland (NCN) via Grant Nos. 2014/14/E/ST3/00256 (AR) and 2014/15/N/ST3/03761 (GR).  
Computations were partly performed using the PL-Grid infrastructure.


\appendix
\section{Transfer-matrix approach in the angular momentum space}
A version of the transfer-matrix approach utilized in this paper is chosen such that the differential equation for transfer matrix can be directly derived from Eqs.\ (\ref{eq:uklRow}) and (\ref{eq:glr}) [see Section IIB] for the wavefunctions, and solved analytically in the absence of trigonal warping ($\nu\equiv{}t'/t_0=0$). The numerical procedure for $\nu\neq{}0$ is  described below. 

First, a~basis set for the transfer matrix is constructed starting from a~general angular-momentum eigenstate in the disk area ($R_{\rm i}<R_{\rm o}$), corresponding to a~given angular-momentum quantum number $l=0,\pm{}1,\pm{}2,\dots$, a~radial part of which can be written as a linear combination of four spinor functions $\phi^{l}\left(r\right)=\sum_{\alpha=1}^{4}a_{\alpha}^{l}\phi_{\alpha}^{l}\left(r\right)$.
The coefficients $\{a_{\alpha}^{l}\}$ are arbitrary amplitudes and $\phi_{\alpha}^{l}\left(r\right)=\left(\phi_{1}^{\alpha,l},\phi_{2}^{\alpha,l},\phi_{3}^{\alpha,l},\phi_{4}^{\alpha,l}\right)^{T}\left(r\right)$
is a normalized spinor function. (The normalization is carried
out in such a way that the total radial current remains constant.)
The wavefunction
$\phi^{l}(r)$ can be represented as 
\begin{equation}
  \label{eq:wf}
  \phi^{l}\left(r\right)=\mathbb{W}^{l}\left(r\right)\boldsymbol{a}^{l},
\end{equation}
where $\mathbb{W}^{l}\left(r\right)$ is the $4\times{}4$ matrix with elements $\left[\,\mathbb{W}^{l}\left(r\right)\,\right]_{m,n}=\phi_{m}^{n,l}\left(r\right)$,
and $\boldsymbol{a}^{l}=\left(\begin{array}{cccc}
a_{1}^{l}, & a_{2}^{l}, & a_{3}^{l}, & a_{4}^{l}\end{array}\right)^{T}$. 

Next, the radial part of the actual wavefunction [see Eq.\ (\ref{eq:cAnM}) in the main text], describing the system in the presence of trigonal warping, is truncated by the linear combination of limited number ($2M+1$) of basis functions, each given by Eq.\ (\ref{eq:wf}), corresponding to angular-momentum quantum numbers $l=-M,\dots,M$. Namely, we define
\begin{equation}
  \label{eq:phir1234}
  \mbox{\boldmath{$\phi$}}\left(r\right)=
  \left[\boldsymbol{\phi}_{1}(r),\boldsymbol{\phi}_{2}(r),
    \boldsymbol{\phi}_{3}(r),\boldsymbol{\phi}_{4}(r)\right]^{T},
\end{equation}
 with
\begin{equation}
  \label{eq:phiimijr}
  \boldsymbol{\phi}_{i}\left(r\right)=
  \sum_{j=1}^4\mathbb{M}\left(i,j;r\right)\boldsymbol{a}_{j}, 
  \ \ \ \ (i=1,\dots,4),
\end{equation}
where $\boldsymbol{a}_{j}=\left[a_{j}^{-M},\ldots,a_{j}^{M}\right]^{T}$,
and the $(2M+1)\times(2M+1)$ matrix $\mathbb{M}\left(i,j;r\right)$ is to be specified later in this Appendix. The relation between wavefunctions at different radii, say $r$ and $R_{\rm i}$, can be expressed within the propagator $\mathbb{U}\left(r,R_{i}\right)$ as follows
\begin{eqnarray}
  \label{eq:prop}
  \boldsymbol{\phi}\left(r\right) & = & \mathbb{U}\left(r,R_{i}\right)\boldsymbol{\phi}\left(R_{i}\right).
\end{eqnarray}
Substituting Eq.\ (\ref{eq:prop}) into Eq.\ (\ref{eq:uklRow}) in the main text, we obtain
\begin{equation}
  \label{eq:rroz}
  \partial_{r}\mathbb{\mathbb{U}}\left(r,R_{i}\right)=
  \mathbb{A}\left(r\right)\mathbb{\mathbb{U}}\left(r,R_{\rm i}\right),
\end{equation}
with the boundary condition 
\begin{equation}
  \label{eq:bcuri}
  \mathbb{\mathbb{U}}\left(R_{\rm i},R_{\rm i}\right)=
  \mathbb{I}_{\left(8M+4\right)\times\left(8M+4\right)},
\end{equation}
where $\mathbb{I}_{N\times{}N}$ denotes the $N\times{}N$ identity matrix. 
The sparse matrix $\mathbb{A}(r)$ in Eq. (\ref{eq:rroz}) has nonzero elements $\left[\,\mathbb{A}(r)\,\right]_{m,n}$ directly following from Eq.\ (\ref{eq:uklRow}). Defining 
\begin{align}
  \label{eq:lmalpm}
  l_m & = M-\left[\left(m-1\right)\mbox{mod}\left(2M+1\right)\right], 
  \nonumber \\
  \alpha_m & = \left\lfloor \left(m-1\right)/\left(2M+1\right)\right\rfloor,
\end{align}
where $\left\lfloor x\right\rfloor$ denotes the largest integer smaller or equal to $x$, one can write down 
\begin{widetext}
\begin{align}
  \label{eq:armn}
  \left[\,\mathbb{A}(r)\,\right]_{m,n} & = 
  \left[
    -g(l_m,r)\delta_{m,n}+i\nu{}t\delta_{m,n-3}+2\nu{}g(l_m\!-\!2,r)\delta_{m,n-6M-6}
  \right]\delta_{\alpha_m,0}
  +
  \left[
      g(l_m\!-\!1,r)\delta_{m,n} - it\delta_{m,n-2M-1}
  \right]\delta_{\alpha_m,1}  
  \nonumber \\
  & + 
  \left[
    g(l_m,r)\delta_{m,n}+i\nu{}t\delta_{m,n+3}-2\nu{}g(l_m\!+\!2,r)\delta_{m,n+2M+4}
  \right]\delta_{\alpha_m,2}  
  -
  \left[
    g(l_m\!+\!1,r)\delta_{m,n}+it\delta_{m,n+6M+3}
  \right]\delta_{\alpha_m,3},  
\end{align}
\end{widetext}
where $1\leqslant{}m,n\leqslant{}8M+4$, the symbol 
$\delta_{\alpha,\beta}$ denotes the Kronecker delta of $\alpha$ and $\beta$, $t=t_\perp/(\hbar{}v_F)$, and $g(l,r)$ is given by Eq.\ (\ref{eq:glr}) in the main text. 

The angular-momentum cutoff $M$ is chosen to be large enough to reach the convergence of the charge-transfer characteristics. We observe that the desired relative precision of $10^{-4}$ requires $M$ growing approximately linearly with the system size $R_{\rm o}-R_{\rm i}$, the magnetic field $B$, and the skew-interlayer hopping $t'$. For instance, in our numerical examples with $t'=0.3\,$eV, the number of modes varies from $2M+1=50$ for $R_{\rm o}-R_{\rm i}=90\,$nm and $B=0$, up to $2M+1=1000$ for $R_{\rm o}-R_{\rm i}=5\,\mu$m and $B=80\,$T. It is also worth to mention, that in the magnetic field $B>0$ efficient computation requires the angular momentum quantum numbers are varied in a~range 
\begin{equation}
  l=-M-\lfloor\Phi_D/\Phi_0\rfloor,\dots,M-\lfloor\Phi_D/\Phi_0\rfloor,
\end{equation}
where $\Phi_D=\pi{}(R_{\rm o}^2-R_{\rm i}^2)B$. 

The numerical integration of Eq.\ (\ref{eq:rroz}), with the boundary condition given by Eq.\ (\ref{eq:bcuri}) and the matrix $\mathbb{A}(r)$ given by Eqs.\ (\ref{eq:lmalpm},\ref{eq:armn}), was carried out by employing the $4$-th order explicit Runge-Kutta method with a~fixed step size \cite{press,istepfoo}. Floating-point arithmetic, with up to $300$ decimal digits, was used to guarantee the numerical stability when inverting the blocks of the resulting transfer matrix for the whole system (see below). 

A procedure, described in the above, brings us to the propagator for the disk area $\mathbb{U}(R_{\rm o},R_{\rm i})$. Writing down the standard mode-matching conditions for wavefunctions in the leads and in the disk area
\begin{align}
  \boldsymbol{\phi}^{\rm lead}\left(R_{\rm o}\right) &= 
  \boldsymbol{\phi}^{\rm sample}\left(R_{\rm o}\right), \nonumber \\
  \boldsymbol{\phi}^{\rm sample}\left(R_{\rm i}\right) &= 
  \boldsymbol{\phi}^{\rm lead}\left(R_{\rm i}\right),
\end{align}
together with Eq.\ (\ref{eq:prop}) for $r=R_{\rm o}$, gives us 
\begin{equation}
  \boldsymbol{\phi}^{\rm lead}\left(R_{\rm o}\right) = 
  \mathbb{\mathbb{U}}\left(R_{\rm o},R_{\rm i}\right)
  \boldsymbol{\phi}^{\rm lead}\left(R_{\rm i}\right).
\end{equation}
In order to find the transfer matrix for the whole system, we choose the wavefunctions in the leads such that
\begin{align}
  \boldsymbol{\phi}^{\rm lead}\left(R_{\rm o}\right) &= 
  \mathbb{M}_{\rm lead}(R_{\rm o})\,\boldsymbol{a}, \nonumber \\
  \boldsymbol{\phi}^{\rm lead}\left(R_{\rm i}\right) &= 
  \mathbb{M}_{\rm lead}(R_{\rm i})\,\boldsymbol{b}, 
\end{align}
where the vector $\boldsymbol{a}$ ($\boldsymbol{b}$) contains $8M+4$ amplitudes for normal modes in the outer (inner) lead. Taking the limit of infinite doping in the leads one can disregard the parameter $\nu$, and write down
\begin{equation}
  \mathbb{M}_{\rm lead}\left(r\right)=
  \mathbb{B}\left(r\right)\otimes\mathbb{I}_{\left(2M+1\right)\times\left(2M+1\right)},
\end{equation}
for $r<R_{\rm i}$ or $r>R_{\rm o}$, where
\begin{equation}
\mathbb{B}\left(r\right)=\frac{1}{\sqrt{r}}\left(\begin{array}{cccc}
1 & 1 & 1 & 1\\
1 & 1 & -1 & -1\\
-1 & 1 & -1 & 1\\
-1 & 1 & 1 & -1
\end{array}\right),
\end{equation}
$\mathbb{A}\otimes\mathbb{B}$ denotes the Kronecker product of the matrices $\mathbb{A}$ and $\mathbb{B}$, and we have further skipped the physically-irrelevant constant phase. In turn, the matrices $\mathbb{M}(i,j;r)$ defining the wavefunction via Eqs.\ (\ref{eq:phir1234}) and (\ref{eq:phiimijr}) can be found as blocks of the matrix $\mathbb{U}(r,R_{\rm i})\mathbb{M}_{\rm lead}(R_{\rm i})$. 

The transfer matrix thus reads
\begin{multline}
  \mathbb{T} =
  \mathbb{M}_{\rm lead}^{-1}\left(R_{\rm o}\right)
  \mathbb{U}\left(R_{\rm o},R_{\rm i}\right)
  \mathbb{M}_{\rm lead}\left(R_{\rm i}\right) \\
   =
  \left(\begin{array}{cc}
      \left(\mathbf{t}^{\dagger}\right)^{-1} 
      & \mathbf{r}'\cdot\left(\mathbf{t}'\right)^{-1} \\
      -\left(\mathbf{t}'\right)^{-1}\cdot\mathbf{r}' 
      & \left(\mathbf{t}'\right)^{-1}
\end{array}\right), \label{eq:transfer}
\end{multline}
where the rightmost equality maps the matrix blocks of $\mathbb{T}$ onto the elements of the scattering matrix: $\mathbf{t}$, $\mathbf{r}$ -- the transmission and reflection matrices for a wavefunction incoming from the inner lead, and $\mathbf{t}'$, $\mathbf{r}'$ -- the transmission and reflection matrices for a wavefunction incoming from the outer lead \cite{Naz}.

\section{
  Wavefunctions for a rectangular sample in uniform magnetic field
}
In this Appendix we present the wavefunctions utilized in Sec.\ IVB to discuss the magnetotransport through a rectangular BLG sample at the charge-neutrality point. The low-energy Hamiltonian has a~general form as given by Eq.\ (\ref{hameffu}) in the main text, but the potential energy now depends only on the $x$-coordinate 
\begin{equation}
  U(x) = \left\{
    \begin{array}{lll}
      U_\infty, & \text{if} & x<0 \text{ or } x>L, \\
      0, & \text{if} & 0<x<L,
    \end{array}
  \right.
\end{equation}
and we choose the Landau gauge $(A_x,A_y)=(0,Bx)$. Subsequently, the solution of the Dirac equation $H\psi=E\psi$, corresponding to a~given transverse wavenumber $k$, can be written as $\psi(x,y)=\phi_k(x)\exp(iky)$. (For the sample width $W$ and the periodic boundary conditions along $y$-direction, we have $k=0,\pm{}2\pi/W,\pm{}4\pi/W,\dots$.)

Using the compact notation: $\psi_{\beta}^{\alpha}\left(x\right)\equiv\psi_{\beta}^{\alpha}$, $\gamma\equiv[\,\gamma_{k}\left(x\right)/16\,]^{2/3}$ [see Eq.\ (\ref{eq:gammak}) in the main text], and $\chi=x\left[k+it\nu/2-x/\left(2l_{B}^{2}\right)\right]$, one can write down a~zero-energy wavefunction for the sample area ($0<x<L$) as a combination of four linearly-independent spinors
\begin{multline}
  \label{phiksamp}
  \phi_k^{\rm sample}(x) = 
  C_1^k \left( \begin{array}{c}
      \psi^{1}_{1} \\ 0 \\ 0 \\ \psi^{4}_{1} 
    \end{array}\right)
  +C_2^k \left( \begin{array}{c}
      \psi^{1}_{2} \\ 0 \\ 0 \\ \psi^{4}_{2} 
    \end{array}\right)
  \\
  +C_3^k \left( \begin{array}{c}
      0 \\ \psi^{2}_{1} \\ \psi^{3}_{1} \\ 0 
    \end{array}\right)
  +C_4^k \left( \begin{array}{c}
      \psi^{2}_{2} \\ 0 \\ \psi^{3}_{2} \\ 0 
    \end{array}\right),
\end{multline}
with $C_1^k,\dots,C_4^k$ being arbitrary coefficients. 
The spinor components in Eq.\ (\ref{phiksamp}) are given by
\begin{eqnarray}
  \psi_{1}^{1} & = & 
  e^{\chi}\left[\,i\tau\text{Ai}'\left(\gamma\right)-(\nu/2)\text{Ai}\left(\gamma\right)\,\right],\nonumber \\
  \psi_{2}^{1} & = & 
  e^{\chi}\left[\,i\tau\text{Bi}'\left(\gamma\right)-(\nu/2)\text{Bi}\left(\gamma\right)\,\right],\nonumber \\
  \psi_{1}^{2} & = & 
  e^{-\chi^{*}}\text{Ai}\left(\gamma^{*}\right),\nonumber \\
  \psi_{2}^{2} & = & 
  e^{-\chi^{*}}\text{Bi}\left(\gamma^{*}\right),\nonumber \\
  \psi_{1}^{3} & = & 
  e^{-\chi^{*}}\left[\,i\tau^{*}\text{Ai}'\left(\gamma^{*}\right)-(\nu/2)\text{Ai}\left(\gamma^{*}\right)\,\right],\nonumber \\
  \psi_{2}^{3} & = & 
  e^{-\chi^{*}}\left[\,i\tau^{*}\text{Bi}'\left(\gamma^{*}\right)-(\nu/2)\text{Bi}\left(\gamma^{*}\right)\,\right],\nonumber \\
  \psi_{1}^{4} & = & 
  e^{\chi}\text{Ai}\left(\gamma\right),\nonumber \\
  \psi_{2}^{4} & = & 
  e^{\chi}\text{Bi}\left(\gamma\right), \label{psi14rec}
\end{eqnarray}
where $\mbox{Ai}\left(z\right)$ and $\mbox{Bi}\left(z\right)$ are
the Airy functions \cite{olv}, and we have further defined 
$\tau=\sqrt[3]{-2i\nu t^{-2}}$.

Remaining details of the mode-matching analysis are same as in Refs.\ \cite{Sny07,Rut14a}. 





\begin{thebibliography}{}
\bibitem{gasymfoo}  
We further notice a~very slow convergence of the actual $G_{\rm BLG}(t'=0)$ obtained from Eqs.\ (\ref{gblg0dsk}--\ref{upsidef}) to the asymptotic form in Eq.\ (\ref{gblgasym}). In particular, for $R_{\rm i}=10\,\l_\perp$ a $5\%$ deviation is still observed for $R_{\rm o}/R_{\rm i}\approx{}1000$. In comparison, for $G_{\rm MLG}$ a~better agreement with the asymptotic form in Eq.\ (\ref{gmlgdisk}) occurs for $R_{\rm o}/R_{\rm i}\geqslant{}38$.
\bibitem{istepfoo}
The integration step needs to be shrunken approximately linearly with the system size and magnetic field, in order to reach the desired relative precision ($10^{-4}$) for the output charge-transfer cumulants. For instance, if $R_{\rm o}-R_{\rm i}=90\,$nm, the step varies from $\Delta{}r=10^{-10}\,$m ($B=0$) down to  $\Delta{}r=10^{-11}\,$m ($B=80\,$T); for $R_{\rm o}-R_{\rm i}=5\,\mu$m and $B=0$ the desired precision is achieved at $\Delta{}r=10^{-8}\,$m.
\end{thebibliography}
\end{document}